\documentclass[fleqn,usenatbib]{mnras}

\usepackage{newtxtext,newtxmath}

\usepackage[T1]{fontenc}

\DeclareRobustCommand{\VAN}[3]{#2}
\let\VANthebibliography\thebibliography
\def\thebibliography{\DeclareRobustCommand{\VAN}[3]{##3}\VANthebibliography}

\usepackage{graphicx}	
\usepackage{amsmath}	
\usepackage{multirow}   
\usepackage{float}
\usepackage{xcolor}

\usepackage{tikz}
\usepackage{amsmath}
\usepackage{bm}
\usepackage{pdfpages}

\title[Day and Night]{Day and Night: Habitability of Tidally Locked Planets with Sporadic Rotation}

\author[Shakespeare \& Steffen]{
Cody J. Shakespeare$^{1}$
and Jason H. Steffen$^{1}$\thanks{E-mail: jason.steffen@unlv.edu}
\\
$^{1}$Department of Physics and Astronomy, University of Nevada, Las Vegas, 4505 S. Maryland Pkwy, Las Vegas 89154, USA\\
}

\date{Accepted XXX. Received YYY; in original form ZZZ}

\pubyear{2023}

\begin{document}
\label{firstpage}
\pagerange{\pageref{firstpage}--\pageref{lastpage}}
\maketitle

\begin{abstract}
Tidally locked worlds provide a unique opportunity for constraining the probable climates of certain exoplanets. They are unique in that few exoplanet spin and obliquity states are known or will be determined in the near future: both of which are critical in modeling climate. A recent study shows the dynamical conditions present in the TRAPPIST-1 system make rotation and large librations of the substellar point possible for these planets, which are usually assumed to be tidally locked. We independently confirm the tendency for planets in TRAPPIST-1-like systems to sporadically transition from tidally locked libration to slow rotation using N-body simulations. We examine the nature and frequency of these spin states to best inform energy balance models which predict the temperature profile of the planet’s surface. Our findings show that tidally locked planets with sporadic rotation are able to be in both long-term persistent states and states with prolonged transient behavior: where frequent transitions between behaviors occur. Quasi-stable spin regimes, where the planet exhibits one spin behavior for up to hundreds of millennia, are likely able to form stable climate systems while the spin behavior is constant. 1D energy balance models show that tidally locked planets with sporadic rotation around M-dwarfs will experience a relatively small change in substellar temperature due to the lower albedo of ice in an infrared dominant stellar spectrum. The exact effects of large changes in temperature profiles on these planets as they rotate require more robust climate models, like 3D global circulation models, to better examine.
\end{abstract}




\section{Introduction}

Tidally locked planets have spin rates synchronous with their orbital rate. This creates a unique situation where, unperturbed, the planet will have a permanent day and night side. The unperturbed scenario assumes that the tidally locked planet does not have neighboring planets that exert orbital changes capable of breaking spin-orbit synchronization. However, many known exoplanet systems have multiple planets in compact orbits close to their star. Furthermore, many of these systems contain planets in mean motion resonance (MMR) with integer ratios of their orbital periods or resonant chains of multiple planets in consecutive MMRs. Planets that orbit close to their stars are typically assumed to be tidally locked due to increased tidal forces and dissipation. In multiplanet systems, these same planets are more susceptible to MMR and resonant chains due to their close proximity with neighbors. Thus, in certain instances, the planets we most often associate with tidal locking are also the most likely to be perturbed out of spin-orbit synchronization. This paper marks the completion of the work presented in \citet{Shake2022}.

Previous works in studying the climates of tidally locked planets that lie within the habitable zone have yielded many insights into how tidal locking impacts habitability. Many studies have shown that with sufficient heat circulation in the atmosphere and/or oceans, these planets may not have a temperature dichotomy as extreme as was once thought between their day and night side.

\citet{Salazar20} examined the impact of substellar continents with varying sizes on Proxima Centauri b. They used the ROCKE-3D General Circulation Model (GCM) with a coupled ocean-atmosphere to examine heat transport, ocean dynamics, and nutrient transport from upwelling in the oceans. \citet{Salazar20} assumes a permanently tidally locked planet, which may be accurate for Proxima Centauri b if there are no neighboring planets in the system. Earlier, \citet{Lewis2018} conducted a similar study examining continent size on Proxima Centauri b using the Global Atmosphere 7.0 configuration of the Met Office Unified Model, another 3D GCM. Their model did not include ocean dynamics but it did examine precipitation. Both \citet{Lewis2018} and \citet{Salazar20} found that substellar continents increase the day-night temperature contrast. \citet{Salazar20} found that heat transport patterns transition when the substellar continent exceeds about $20\%$ of the planet's surface area.

\citet{Yan2020} investigated the presence of hurricanes on tidally locked planets using the global Community Atmosphere Model version 4. They used a fixed surface temperature profile derived from previous GCMs of tidally locked planets. \citet{Yan2020} finds that hurricanes can occur on tidally locked planets, sometimes with high frequency, depending on semi-major axis and atmospheric composition. They also include an impressive supplementary video of the hurricanes in their appendix.

Many previous climate studies of tidally locked planets assume the planets are perfectly tidally locked. However, many tidally locked planets are in multiplanet systems where the planets will experience orbital forcing from their neighbors. It should be noted, however, that single-planet systems, multiplanet systems with large orbital spacing, or certain multiplanet architectures may be sufficiently described as near perfectly tidally locked.

The recent paper by \citet{Chen2023} models the climate impact of tidally locked planets with sporadic rotation using an N-Rigid-Body integrator, GRIT, and the 3D GCM ExoPlaSim \citep{Paradise2022}. They use TRAPPIST-1 e and f as examples of planets that may be tidally locked with sporadic rotation (TLSR). \citet{Chen2023}'s results show that certain spin histories of the TLSR spin state can cause permanent snowball states, especially in the outer habitable zone.

The nature of these not-quite-synchronous spin states is different from both the tidally locked state and the rotating spin states observed in the solar system's terrestrial planets. The primary difference is how the spin behavior and direction, with respect to the star, can change on the order of years. \citet{Vinson2019}, who conducted a similar analysis as this paper on spin behavior, describe this unique spin state as 'chaotic'. We classify this spin state, generally, as 'tidally locked with sporadic rotation' (TLSR). The TLSR spin state can be broadly described as chaotic. However, we opt for a more descriptive name, TLSR, to encompass this complex spin state. 

The TLSR spin state is unique in that the spin behavior is often not consistently tidally locked nor is it consistently rotating. Instead, the planet may suddenly switch between spin behaviors that have lasted for only a few years or up to hundreds of millennia. The spin behavior can occasionally be tidally locked with small or large librations in the longitude of the substellar point. The planet may flip between stable tidally locked positions by spinning $180^{\circ}$ so that the previous substellar longitude is now located at the new antistellar point, and vice versa. The planet may also spin with respect to the star, having many consecutive full rotations. The spin direction can also change, causing prograde and retrograde spins. Although to be clear, while the planet may change spin direction with respect to the star and the substellar point, the planet's intrinsic spin compared to an outside observer does not change direction. Lastly, the planet may exhibit prolonged transient behavior (PTB), where frequent and successive transitions between behaviors occur for an extended period of time without settling. See Section \ref{sec:Analyses} for full descriptions of the four common types of spin behavior we classify as 'spin regimes'.

The often inconsistent behavior of the TLSR spin state is an interesting prospect for habitability and climate as the temperature fluctuations of any point on the planet will be extreme, perhaps excluding around the poles. However, the persistent nature of the different spin regimes is another factor to consider. Such a planet may not have a single, persistent global climate as it is described for Earth, but it may have multiple climate states that are determined by the present and past spin regimes as well as the response time of the climate system to a change in spin. One can imagine many complicated situations that these planets would experience in the TLSR state.


An illustrative example would be a planet that was previously tidally locked for a long period of time, hundreds to thousands of years, whatever is long enough that the climate has settled into a stable state. Such a planet in the habitable zone around a TRAPPIST-1-like star could have an orbital period of around 4-12 Earth days -- the approximate orbital periods of T-1d and T-1g, respectively. Due to the TLSR spin state, this planet may, rather abruptly, start to rotate, albeit slowly -- on the order of one rotation every few Earth years. The previous night side of the planet, which had not seen starlight for many Earth years, will now suddenly be subjected to variable heat with a day-night cycle lasting a few years. The day side would receive a similar abrupt change and the climate state that prevailed for centuries would suddenly be a spinning engine with momentum but spark plugs that now fire out-of-sync with the pistons. In this analogy, the spark plugs and the subsequent ignition of fuel correspond to the input of energy from starlight. The response of ocean currents, prevailing winds, and weather patterns may be quite dramatic. However, the complicated details of the impact of such events, which almost certainly do occur, would be difficult to model, especially without an accurate geography to apply it to.

\section{Methods}
\label{sec:methods}

\begin{figure}
    \centering
    \resizebox{\columnwidth}{!}{%
    \begin{tikzpicture}
    \usetikzlibrary{calc}

    \newcommand{\tikzAngleOfLine}{\tikz@AngleOfLine}
      \def\tikz@AngleOfLine(#1)(#2)#3{%
      \pgfmathanglebetweenpoints{%
        \pgfpointanchor{#1}{center}}{%
        \pgfpointanchor{#2}{center}}
      \pgfmathsetmacro{#3}{\pgfmathresult}%
      }

    \draw (-5,0) node[anchor=north]{\LARGE\textbf I} --
    (5,0) node[anchor=north]{\LARGE\textbf F}
    (-3,0) node{}--
    (3,3) node[anchor=west]{\LARGE\textbf P} --
    (2,0) node{}
    (-3,0) node{}--
    (-4.5,-0.75) node[anchor=north]{\LARGE\textbf S};
    \draw [color=black, fill=black] (-3,0) circle(0.04)
    (3,3) circle(0.04);

    \coordinate (A) at (-3,0);
    \coordinate (B) at (2,0);
    \coordinate (P) at (3,3);
    \coordinate (F) at (5,0);
    \coordinate (I) at (-5,0);

     \tikzAngleOfLine(I)(F){\AngleStart}
        \tikzAngleOfLine(A)(P){\AngleEnd}
        \draw[black, ->] (A)+(\AngleStart:1.5cm) arc (\AngleStart:\AngleEnd:1.5cm);
      \node at ($(A)+({(\AngleStart+\AngleEnd)/2}:1 cm)$) {\LARGE$\textbf M$};

     \tikzAngleOfLine(P)(A){\AngleStart}
        \tikzAngleOfLine(P)(B){\AngleEnd}
        \draw[black, ->] (P)+(\AngleStart:1.3cm) arc (\AngleStart:\AngleEnd:1.3cm);
      \node at ($(P)+({(\AngleStart+\AngleEnd)/2}:1.1cm)$) {\LARGE$\boldsymbol \gamma$};
      \node[anchor=north west] at ($(P)+(\AngleEnd:1cm)$){\LARGE\textbf J};
    \draw [color=black, fill=black] ($(P)+(\AngleEnd:1cm)$) circle(0.04);

     \tikzAngleOfLine(B)(F){\AngleStart}
        \tikzAngleOfLine(B)(P){\AngleEnd}
        \draw[black, ->] (B)+(\AngleStart:1cm) arc (\AngleStart:\AngleEnd:1cm);
      \node at ($(B)+({(\AngleStart+\AngleEnd)/2}:0.7cm)$) {\LARGE$\boldsymbol \theta$};


    \draw [rotate around={-20:(3,3)}] (P) ellipse (0.5cm and 1cm);
    \end{tikzpicture}
    }
        \caption{Illustration of the substellar longitude, $\gamma$, and its relation to other angles and lines. Line IF is an inertial line and line PS is the line connecting the center of the star and the planet. The ellipse surrounding point P represents the planet's geometry as a triaxial ellipsoid. This figure is a recreation of Figure 2 of \citet{Gold1966} when a planet is near 1:1 synchronous rotation.}
    \label{fig:gamma}
\end{figure}
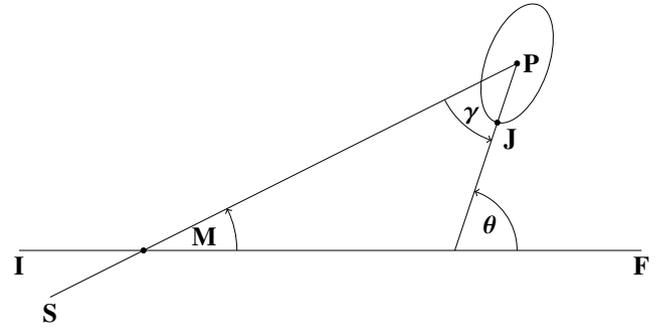

All tidally locked bodies experience some degree of libration, even absent any third bodies. For the sake of clarity, let us define a planet as tidally locked when its libration is less than $2$ radians, or about $114^{\circ}$. Direct starlight does not reach the antistellar point, opposite the substellar point, when the libration is less than $\frac{\pi}{2}$ rad. However, with hindsight, $2$ radians is a better cutoff for when spin behavior changes in our spin integrations.

The substellar longitude is defined by the difference between two angles. $\theta$ is the angle between the long axis of the planet and a stationary inertial line \citep{Gold1966}. The substellar longitude is then defined as $\gamma=\theta - M$ where $M$ is the mean anomaly (see Figure \ref{fig:gamma}). We use the equation of motion derived by \citet{Gold1966}, \citet{Gold1968}, and \citet{Murray1999} which was simplified in its presentation by \citet{Vinson2019}. The equation of motion for the spin of a planet near 1:1 synchronous rotation under tidal forces is given by
\begin{equation}
    \ddot \gamma + sgn[H(e)]\frac{1}{2}\omega_s^2sin(2\gamma)+\dot n = 0,
    \label{eq:gam}
\end{equation}
where
\begin{equation}
    H(e)=1-\frac{5}{2}e^2+\frac{13}{16}e^4
\end{equation}
and
\begin{equation}
    \omega_s^2=3n^2\left(\frac{B-A}{C}\right)|H(e)|.
\end{equation}
The term $\frac{B-A}{C}$ is the triaxiality factor of the three principal moments of inertia where $C$ is the moment about the spin axis, $A$ is the moment about the long axis, and $B>A$. Note that the long axis is aligned with the line PJ in Figure \ref{fig:gamma} and the planet's shape is that of a triaxial ellipsoid.

The equation we use for constant time lag tidal dissipation was derived by \citet{Hut1981} and \citet{Eggleton1998}, later presented by \citet{Hansen2010}, and again tidied up by \citet{Vinson2019}. The equation for constant time lag tidal dissipation is
\begin{equation}
    \ddot \gamma = -\frac{15}{2}\dot \gamma \frac{M_*}{m}\left(\frac{R}{a}\right)^6M_*R^2\sigma,
\end{equation}
where
\begin{equation}
    \sigma=\frac{1}{2Q}\left(\frac{G}{\Omega R^5}\right)
\end{equation}
and $Q$ is the tidal parameter. We follow \citet{Vinson2019} in using a tidal parameter of $Q=10$ and $Q=100$. The former is similar to the Earth and is consistent with planets that have thick atmospheres or Earth-sized oceans. A tidal parameter of $Q=100$ corresponds to a planet with less liquid water, although water may be present in ice sheets. These tidal parameters are consistent with our later assumptions of an ocean covered planet in our energy balance models (EBMs). The assumption of significant water is necessary to examine the maximum bistability of these planets as albedo would not change significantly on a desiccated planet as it does for water-rich planets.

Adding tidal dissipation, the full equation of motion for spin is
\begin{equation}
    \ddot \gamma + \frac{1}{2}\omega_s^2sin(2\gamma)+\dot n + \epsilon \dot \gamma = 0,
    \label{eq:EOM}
\end{equation}
where some constants are simplified by \citet{Vinson2019} in
\begin{equation}
    \epsilon=\frac{15}{2}\frac{M_*}{m}\left(\frac{R}{a}\right)^6M_*R^2\sigma.
\end{equation}

We must retrieve a continuous equation for the eccentricity and mean motion evolution of the planet to integrate Equation \ref{eq:EOM}.  Thus, we use cubic spline interpolation on the eccentricity and mean motion history given by N-body simulations which output the parameters every 1/10th of the target planet's orbital period. To save computation resources, we use overlapping cubic splines, with an overlap of 100 data points on each side of the spline. This results in an error smaller than machine precision on timescales of 1,000 years.

We use the final N-body simulations from \citet{MacDonald2018}. These simulations represent stable systems similar to TRAPPIST-1 formed through short-scale migration, excluding the outermost planet. The outermost planet, TRAPPIST-1h, was excluded at the time due to poor constraints on its mass \citep{Gillon2017}. The low mass of TRAPPIST-1h determined in later studies \citep{Agol2021} means this exclusion has little impact on the results. Furthermore, these are not best-fit systems but instead span a range of plausible systems similar to TRAPPIST-1. Thus, these N-body simulations are better suited for illustrating the range of behaviors that TLSR planets may exhibit, as opposed to representing TRAPPIST-1 specifically. With that said, the amplitude and frequency of eccentricity and mean motion variations in our N-body simulations are similar to the best-fit analyses of \citet{Vinson2019}'s N-body models in \citet{Tamayo2017}. The resulting spin histories are similar to the best-fit analyses in \citet{Vinson2019} in terms of average periods and types of behavior demonstrated. From here on, we refer to our N-body simulations and spin integrations as Planets 2-4 to distinguish between mentions of TRAPPIST-1 specifically and the planets we simulate.

The integration of Equation \ref{eq:EOM} spans 10 million orbits of each planet over which the system remains stable. The migration simulations conducted by \citet{MacDonald2018} for these systems included a stability assessment of about $10^8$ days after migration. Any of the systems which do not pass \citet{MacDonald2018}'s stability check or remain stable during our integration are discarded. Stability is defined by the absence of planet scattering to long-period orbits (>20 Earth days) or orbits with high eccentricity (>0.999). The outputs of spin integration are the substellar longitude, $\gamma$, and the spin rate, $\dot \gamma$.

\begin{figure}
    \centering
    \includegraphics[width=\linewidth]{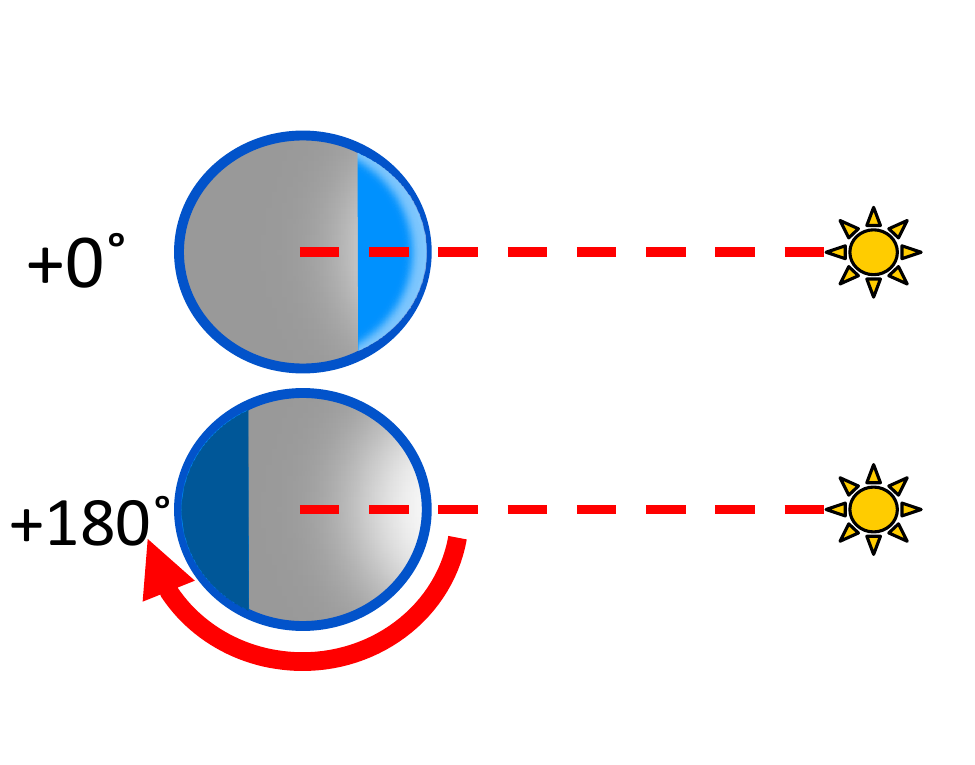}
    \caption{Illustration of albedo rotation in the modified HEXTOR EBM. Each orientation corresponds to an independent EBM. The $180^{\circ}$ EBM uses a constant surface albedo profile taken from the $0^\circ$ EBM.}
    \label{fig:rotate}
\end{figure}

\subsection{Spin Analyses}
\label{sec:Analyses}

We conduct an analysis of the spin integrations by determining average spin/libration periods and examining the nature of continuous spin behaviors, what we call spin regimes. The spin results contain persistent states (classified as short, medium, long, and quasi-stable, in order of increasing length) where the planets may either continuously spin or librate in a tidally locked configuration for hundreds to thousands of years. Determining the most common frequency and duration of these persistent states is important in understanding the stable climates which may form and persist for long periods of time before abruptly ending. Examining persistent states also enables us to measure the relative frequency of PTBs, where the planet may not enter a consistent climate state for long periods of time.

A continuous state, or regime, is defined by a single spin behavior: tidally locked at zero (T.L. Zero), tidally locked at pi (T.L. Pi), and full rotation (Spin/Spinning). These first three regimes are intended to represent consistent spin behaviors that may develop a stable climate system if given enough time. Although, the lower limit of these regimes likely do not afford enough time to enter a new stable system. T.L. Zero and T.L. Pi are described by continuous librations of $\gamma$ at either zero or pi with amplitudes less than 2 radians or $\sim 114^{\circ}$. Spin is described by consecutive rotations in either the clockwise or counterclockwise directions without any interruption; interruptions include a reversal of rotation direction. We define the minimum length for these three regimes as $10$ yrs. The $10$ yr limit is subjective, however, it is useful for simplicity and to avoid classifying short transitionary behavior that is irregular and may be best classified as PTB. The $10$ yr limit also prevents a large fraction of time spent in a regime from being attributed to frequent but short-lived time periods of behavior. 

The prolonged transient behavior (PTB) regime is defined as any period of time where the spin behavior cannot be classified under any of the other regimes and has a minimum length of $0.2$ yrs. The PTB regime is intended to capture prolonged states with frequent transitions between spin behavior. To this end, the PTB lower limit roughly prevents small periods of time between two abutting regimes from being classified as PTB. Accepting the defined limits, examination by eye has found no erroneous classifications (e.g. high amplitude libration being misclassified as PTB, spinning, or the wrong center of libration), although a rare few may exist. Regardless of the ambiguities, the frequency of such regimes and the fraction of the total time that these regimes constitute is interesting.

The four spin regimes are components of the broader TLSR spin state. The TLSR spin state describes a planet that exhibits many spin regimes over its history. The TLSR spin state is distinct from the gradual one-way transition from rotating to tidally locked that planets may experience due to tidal dampening over long timescales. For example, a planet that gradually slows its spin over billions of years until it is tidally locked is not in the TLSR spin state. This planet would be said to have been in a rotating spin state before gradually entering the tidally locked spin state. Although, it is feasible for such a planet to experience a TLSR spin state in its transition from its rotating spin state to its tidally locked spin state.

We define four boundaries that distinguish between regimes of different lengths. Regimes that last less than 100 Earth years are classified as short. Regimes that last between 100 and 500 years are classified as medium. 'Long' regimes last between 500 and 900 years. The longest regimes are classified as quasi-stable and last for 900 years or more. Quasi-stable regimes have the largest range of duration with the longest spanning hundreds of thousands of years.

\begin{figure*}
    \centering
    \makebox[\textwidth]{\includegraphics[width=\textwidth]{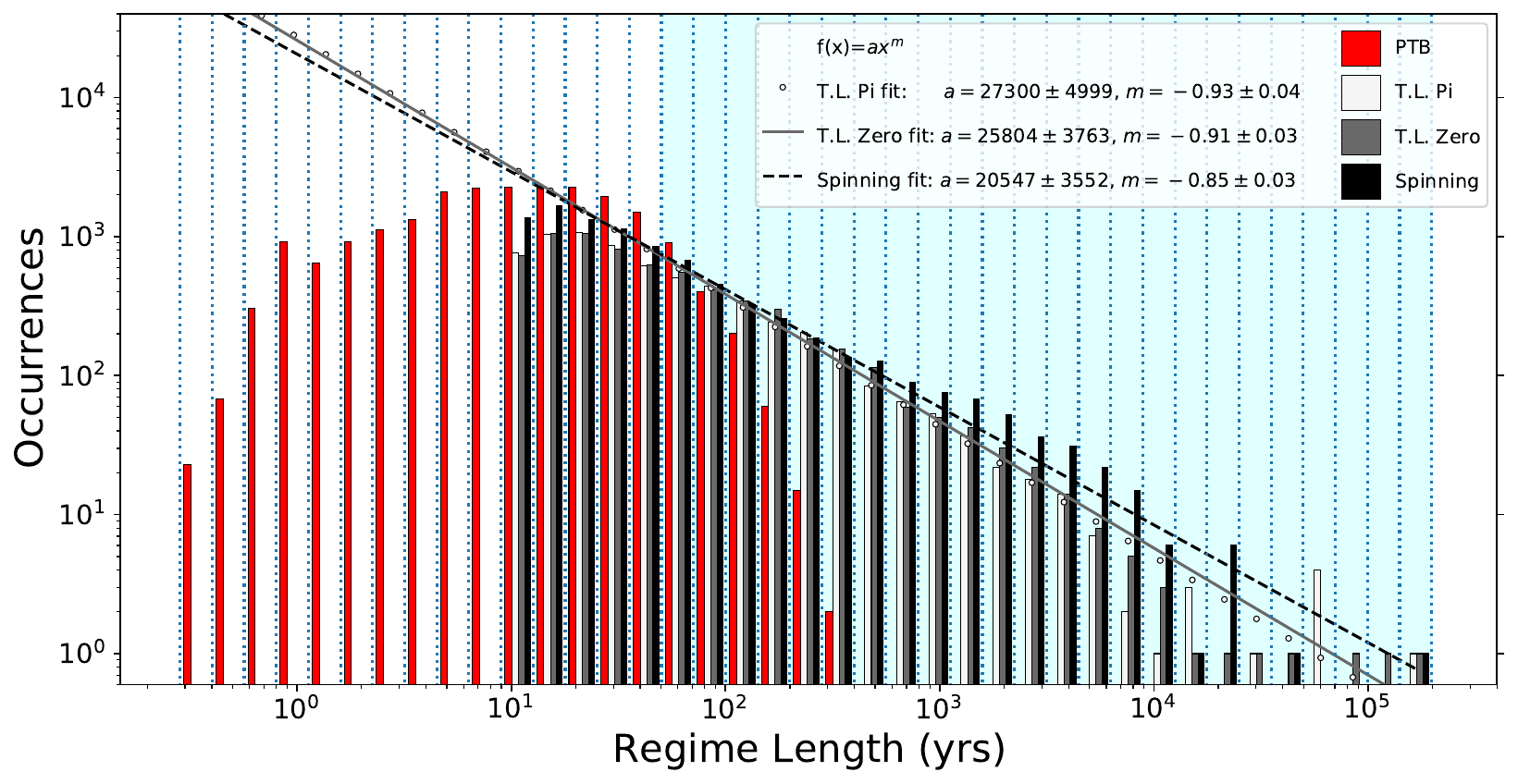}}
    \caption{Histogram of the combined regime occurrences over regime length for all simulations which exhibit the TLSR spin state. The bin boundaries of the log-log histogram also use a log scale and are shown with dotted blue lines. The blue shaded area defines the 'well-behaved' sample data used to find the best-fit power laws for the probability density functions of the three regime types.}
    \label{fig:HistMain}
\end{figure*}

\subsection{Energy Balance Models}
\label{EBM}

We use a modified version of the latest release of the 1D EBM HEXTOR \citep{Haqq2022} to model the climate of a previously tidally locked planet that has rotated $180^{\circ}$ (see Figure \ref{fig:rotate}). The latest release of HEXTOR \citet{Haqq2022} includes the option for a coordinate transform of latitude in a 1D energy balance model. This reorients the latitude's northern pole to the substellar point and the south pole to the antistellar point, a $90^{\circ}$ rotation of the coordinate system. Thus, the east-west coordinate is now latitudinal in nature. Having latitude centered at the substellar point makes use of the fact that stellar insolation on a tidally locked planet's surface is only a function of distance from the substellar point. However, \citet{Haqq2022} refer to east-west degrees as substellar longitude to avoid confusing the reader. We also refer to east-west degrees as substellar longitude to remain consistent with \citet{Haqq2022}.

HEXTOR models using the coordinate transform produces temperature profiles largely consistent with more complex 3D climate models of tidally locked planets that have poor heat transport \citep{Haqq2022}. We use the Hab1 TRAPPIST-1 Habitable Atmosphere Intercomparison (THAI) Protocol \citep{Fauchez2020} for planets c, d, e, and f. The THAI Protocol Hab1 assumes a planet entirely covered in oceans. The surface albedo of Hab1 assumes a stellar spectrum similar to TRAPPIST-1 with a land albedo of 0.3, an ice albedo of 0.25, and an ocean albedo of 0.06. The Hab1 atmosphere is similar to present-Earth with 1 bar $N_2$ and 400 ppm $CO_2$ \citep{Fauchez2020}. We will refer to these EBMs using TRAPPIST-1c - TRAPPIST-1f accordingly as our EBMs use observational measurements of orbital and planetary parameters from \citet{Gillon2017}.

We modify the model to examine the simplified, extreme case where the surface albedo of the planet remains constant over the first half-rotation. Our modifications to the model are solely intended to allow us to rotate the albedo profile from a previous EBM simulation and to then hold that albedo profile constant until the EBM reaches a new equilibrium. While the surface albedo is held constant, the top of atmosphere albedo (TOA) may change due to changes in greenhouse effects modeled in the atmosphere. Thus, we assume no ice melts or forms during rotation. In other words, the night side ice sheets that formed in the tidally locked regime remain after rotation and no new ice forms. This configuration examines how the albedo dichotomy of a tidally locked water planet affects the climate as it begins to spin or when the planet makes a $180^{\circ}$ flip to a new tidally locked position. The goal of this model is to determine the change in surface temperature at each longitude, the change in the substellar temperature ($\Delta T_{sub}$) due to ice sheets, and whether the increase in albedo from ice is likely to force the planet into a snowball state.

The energy balance model reaches an equilibrium in the initial tidally locked configuration. The final surface albedo is then flipped $180^{\circ}$ for the next EBM and the surface albedo is held constant until a new equilibrium is reached. This mimics the simplified case where no ice melts or forms during the $180^{\circ}$ rotation. More precisely, this assumes that any melting of ice sheets during rotation would have an insignificant impact on the albedo during the first half-rotation. While a simplification, this method produces an upper bound (ice-free substellar point) and lower bound (entirely ice covered day-side) of the surface temperature at the substellar point of a planet with poor heat transport. 

Spin periods with lengths on the order of Earth years or more (see Section \ref{sec:SpinResults}) require a large thermal inertia for any point of the surface to contain remnant irradiation history. In other words, with the climate model we use, there is more than enough time for the atmosphere to heat or cool to the temperature defined by the longitude's current stellar irradiation. Thus, we do not rotate the final temperature profile from the $0^\circ$ orientation as it would be an inaccurate approximation: even more so if we consider hysteresis \citep{Haqq2022}. Accordingly, we set the initial surface temperature to 300 K for both orientations. Ample time for temperature change is evident in the Earth's daily and seasonal surface temperature change. Deep oceans or an atmosphere with a higher specific heat could make thermal inertia more significant. The specific heat of the atmosphere can be impacted by surface pressure, chemical composition, and other factors. Accounting for these impacts would also benefit from modeling ocean circulation and heat transport, which is beyond the scope of this paper.

\section{Results}
\label{sec:results}

This section examines the results of our spin integrations and analyses in subsection \ref{sec:SpinResults} for planetary systems similar to TRAPPIST-1. Subsection \ref{sec:EBMresults} examines the results of our modified HEXTOR EBMs for TRAPPIST-1c, d, e, and f.

\subsection{Spin Results}
\label{sec:SpinResults}
The spin behavior can be defined most basically through the period of libration and spin. For any single planet, the period of libration (T.L. Pi and T.L. Zero) and the period of spin are generally the same (see Table \textcolor{blue}{D2}, \textcolor{blue}{D4}, and \textcolor{blue}{D6}). The period is important in determining how quickly surface temperatures change during major excursions from the tidally locked orientation. Major excursions could be defined by large librations and/or full rotation.

The most important determination, at this time, may be characterizing the length of spin regimes and comparing the frequency of long-term and short-term regimes. Using the method described in Section \ref{sec:Analyses}, we divide and sort the spin results by behavior. The abbreviated compilation of period and regime results are shown in Table \textcolor{blue}{D1}-\textcolor{blue}{D6}. The frequency distribution of regimes per regime length, and the fitted probability density function, is shown in Figure \ref{fig:HistMain}. The distribution of spin behavior is shown in terms of the time elapsed in each regime type, or regime length, compared to the total time of integration in Figure \ref{fig:pie}.

\subsubsection{Overall Regime Results}

Figure \ref{fig:HistMain} shows a log-log histogram of regime counts per regime length of the four different regime types. The bin width is also defined on a log scale so the bins appear as a constant width, denoted by the dotted blue lines. The PTB regime shows a roughly symmetrical spread, on this scale, centered around 10 yrs. Since the left side of the PTB regime distribution clearly tapers downward, we can be confident that the 0.2 yr lower limit we set for the PTB regime does not impact results. The nearly constant slope of the other three regimes indicates that the frequency distributions of these regimes are likely described by a power law. We use the Python package scipy.optimize to fit these three probability density functions using the least squares method weighted by the square root of the bin size. The three power laws, their best-fit parameters, and their errors are reported in the legend of Figure \ref{fig:HistMain}. The blue-shaded region of the figure represents the data deemed as 'well-behaved' for the fit. Shorter regime lengths must be excluded as their occurrences are high and may be adversely affected by our minimum regime limit of 10 yrs for the Spinning, T.L. Zero, and T.L. Pi regimes. See Figure \textcolor{blue}{B1} in Appendix \textcolor{blue}{B} for the effect of different choices of data cuts on the power law best-fits.

The length of time required for a planet in the TLSR spin state to produce a regime of a desired, longer length may be useful for future studies. Future studies may want to examine the impacts of the TLSR spin state without the need of preparing and conducting N-body simulations and/or spin integrations. To this end, we derive Equation \ref{eq:SimLength} to calculate the total simulation time, $S$, needed for regimes greater than the desired length $L$ to occur with the predefined probability $P$. The minimum regime length, $R$, is defined as 10 yrs from Section \ref{sec:Analyses} and $m$ is the exponent of the power law, or the slope of the power law on a log-log scale, shown in Figure \ref{fig:HistMain}. See Appendix \textcolor{blue}{D} for the derivation of Equation \ref{eq:SimLength}.
\begin{equation}
    S=\left( \frac{L^{m+1}-PR^{m+1}}{1-P} \right)^{\frac{1}{m+1}}.
    \label{eq:SimLength}
\end{equation}

One question we can now answer, which is also a useful demonstration of Equation \ref{eq:SimLength}, is: what is the simulation length, S, needed if we want to observe regime lengths greater than 1 million years? For the probability of occurrence, let's use roughly the same probability for us to observe regimes greater than 100 kyr over the course of a 200 kyr simulation -- similar to the length of our spin integrations for planet 5. For the Spinning regime, with $m=-0.85$, we calculate this probability at about 12.8\%. So, we set $P(L>1Myr)=0.128$. Using Equation \ref{eq:SimLength}, we calculate that the simulation length needed for a probability of 12.8\% is about 2.1 million years.

The overall regime length averages are shown in Table \ref{tab:avgReg}. These overall averages combine results from planets 2-5 for all simulations which exhibit the TLSR spin state. Similar data tables which show the results of individual simulations and planets can be found in Appendix \ref{sec:appTAB}. All of the data tables are separated with horizontal lines by regime type. Mean values and standard deviation are heavily influenced by the longest of the quasi-stable regimes. The averages in Table \ref{tab:avgReg}, as well as the best-fit results in Figure \ref{fig:HistMain}, show that T.L. Zero and T.L. Pi are more similar to each other than to the Spin regime. The similarity should be expected as the two T.L. regimes are mirrored versions of each other and they are both stable positions. The only difference is that all simulations begin in the T.L. Zero position, so T.L. Zero regimes are accordingly more common overall.

The percentage of time spent in quasi-stable regimes lends insight into how often consistent climate systems may occur. If we calculate the percentage of time spent in quasi-stable regimes for each individual simulation, then the mean percentage is about 29.2\%. In other words, here we have averaged the percentage for quasi-stable regimes seen in Figure \ref{fig:pie}, but for all simulations in the TLSR spin state. The median percentage is nearly the same, at 28.8\%, with the 1st and 3rd quartile at 11.6\% and 45.3\%, respectively. We also sum the total time of all the simulations in the TLSR spin state and find the percentage of total time spent in quasi-stable regimes at 55.4\%. The prevalence of quasi-stable regimes supports the importance of determining the impacts of TLSR spin states. Planets that spend longer periods of time with a consistent spin behavior are more likely to enter a persistent climate state.

\subsubsection{Spin Histories}

\begin{figure*}
    \centering
    \makebox[\textwidth]{\includegraphics[width=\textwidth]{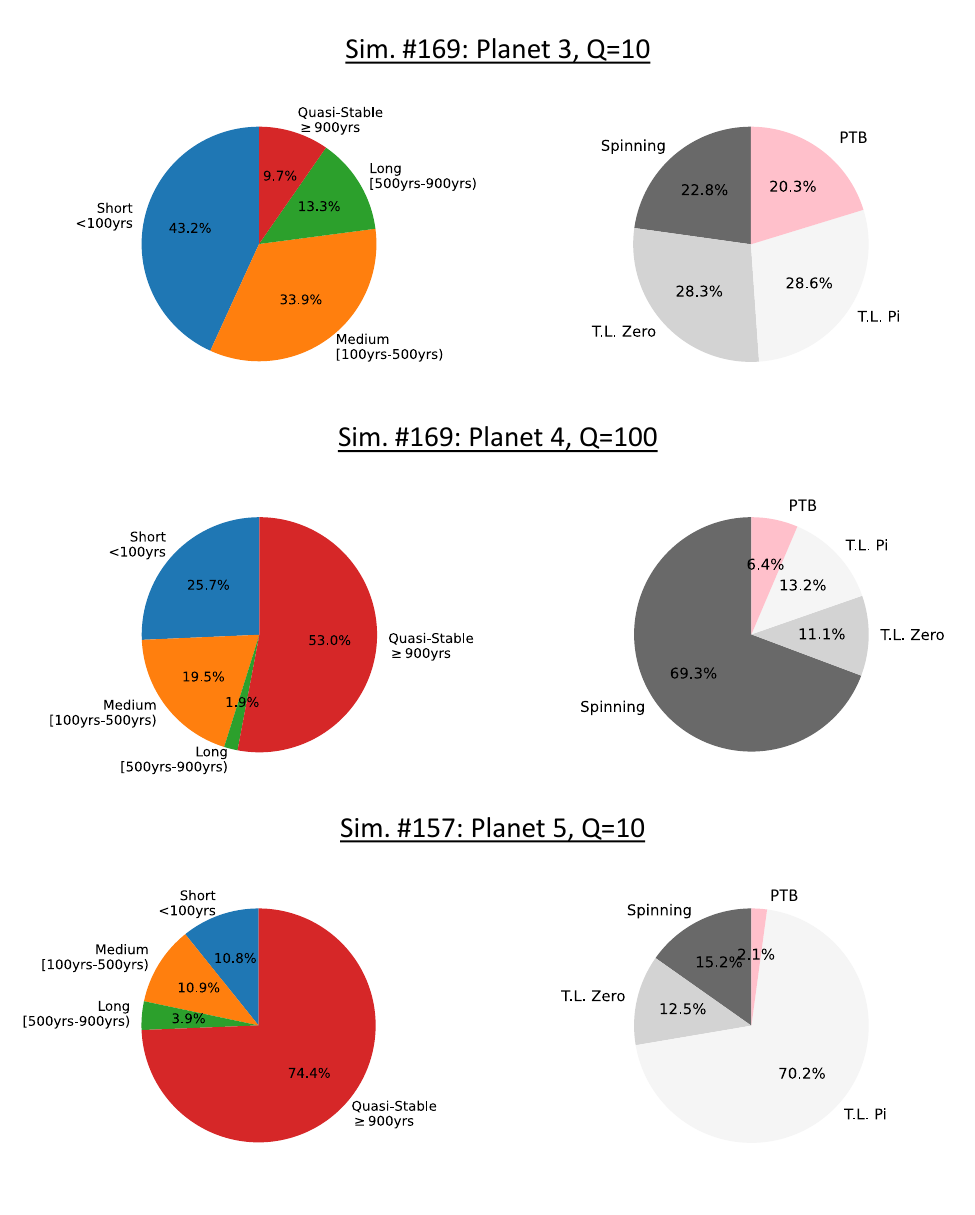}}
    \caption{Percent of total integration time for each regime type or
regime length for select simulations. Columns differ by regimes sorted by length and type. Rows differ by individual spin integrations.}
    \label{fig:pie}
\end{figure*}

Specific spin histories and regime distributions can differ significantly by the planet, the system's orbital evolution, and Q-value. In particular, spin histories are sometimes dominated by a single regime type or regime length (bottom two of Figure \ref{fig:pie}). Other spin histories have a relatively even distribution over all regime types and regime lengths (top of Figure \ref{fig:pie}). It is important to remember these results are not representative of TRAPPIST-1 specifically. Our results are illustrative of the different types of spin behaviors possible which can have varying impacts on climate.

Our spin results generally agree with the previous study of \citet{Vinson2019}, despite our N-body simulations from \citet{MacDonald2018} covering a wide range of TRAPPIST-1-like systems. In the following discussions, we identify simulations that exhibit the TLSR spin state over the course of the spin integration. In other words, a planet that remains in a single regime does not exhibit the qualities of the TLSR spin state over the course of spin integration and we exclude such planets in much of our discussion and analyses of TLSR planets. All of the simulations without TLSR qualities remain tidally locked in our spin integrations. Overall, 28 of the 80 total simulated planets exhibit the TLSR spin state over the time-span integrated. Four simulated planetary systems exhibit the TLSR spin state in all planets of their system: number 152 with Q=10, 152 with Q=100, 169 with Q=10, and 169 with Q=100. These four simulated systems comprise 16 out of the 28 planets in the TLSR spin state. Three simulated systems had only one planet in the TLSR spin state: number 17 with Q=10, 17 with Q=100, and 182 with Q=100. Planet 4 is the most likely planet position to exhibit the TLSR spin state. The remainder of this subsection will explore some noteworthy results.

In simulation number 157 with Q=100, planet 5 (Table \textcolor{blue}{D1}) begins spinning immediately and continues spinning for about 198,000 years. Over time, tidal dissipation slows the spin rate and all regime behaviors occur sporadically. This indicates that the simulation began at a time when the planet's orbit was changing at a faster rate than most scenarios. The initial conditions, primarily the initial spin rate being set equal to the initial orbital rate, play a large factor here. However, it is an example of how orbital forcing on planet spin can vary in magnitude. For the same simulation and planet, but with Q=10 (see Table \textcolor{blue}{D1} and bottom of Figure \ref{fig:pie}), the planet slows its rotation over about 24,000 years. This is expected with a smaller Q-value that corresponds to stronger tidal dissipation. After slowing rotation, the Q=10 planet experiences all regime behaviors sporadically and then enters the T.L. Pi regime for about 175,000 years. This again comes to an end with sporadic regime behaviors.

\begin{table}
	\centering
	\caption{The average lengths for the four regime types over all planets and simulations that exhibit the TLSR spin state. All values are in years.}
	\label{tab:avgReg}
    \begin{tabular}{lrr}
        \\\\
        \hline
        \multirow{5}{*}{Spin}&Mean Len.&184.6\\
        &Std. Dev.&2293.0\\
        &Med. Len.&25.8\\
        &1st Quartile&15.2\\
        &3rd Quartile&59.3\\
        \hline
        \multirow{5}{*}{T.L. Zero}&Mean Len.&196.7\\
        &Std. Dev.&2773.4\\
        &Med. Len.&30.4\\
        &1st Quartile&17.1\\
        &3rd Quartile&76.1\\
        \hline
        \multirow{5}{*}{T.L. Pi}&Mean Len.&188.2\\
        &Std. Dev.&2741.6\\
        &Med. Len.&29.4\\
        &1st Quartile&17.0\\
        &3rd Quartile&73.4\\
        \hline
        \multirow{5}{*}{PTB}&Mean Len.&18.2\\
        &Std. Dev.&22.2\\
        &Med. Len.&10.9\\
        &1st Quartile&4.5\\
        &3rd Quartile&23.9\\
        \hline
    \end{tabular}
\end{table}

Again in simulation number 157, planet 4 experiences a unique spin history that shows how long-term regimes can abruptly end. Figure \textcolor{blue}{B2} shows the spin history of simulation 157, planet 4 for Q=10 and Q=100 (full Q=100 results of planet 4 shown in Figure \textcolor{blue}{D3}). This simulated planet shows two of only three examples where a very long initial regime in T.L. Zero ends and sporadic regime behaviors begin. The initial tidally locked regime lasts about 151,000 years. Note that solid, vertical blue lines are indicative of the Spinning or PTB regimes. They appear solid due to the scale of the image, however, the blue lines are wrapping from $-\pi$ to $\pi$. Any apparent overlap in regime classification is an artifact of low resolution; classified regimes never overlap. For Q=10, the initial T.L. Zero regime lasts 128,000 years, as opposed to about 151,000 years when Q=100. This is a bit counter-intuitive as a Q-value of 10 has stronger tidal dissipation and, thus, produces more resistance to perturbations. The explanation here depends on the rotation direction the planet was in when it received orbital forcing. Seeing as the Q=100 spin integration experiences a larger amplitude of libration due to decreased dissipation, it has a larger angular momentum that must be counteracted with orbital forcing on the spin when it is applied in the opposite rotational direction. The orbital forcing may also be applied at an approximate frequency with an opposite phase (phase shifted by $\pi$ radians) with the planet's libration frequency, resulting in the same situation occurring many times over. A good analogy is trying to push a person on a swing. Adding force parallel to the swinger's velocity results in a larger libration amplitude. If the force is applied at a time when the swinger's velocity is anti-parallel with the force, the result is a decrease in velocity and libration amplitude. It is important to also note that neither of the behavior changes seen in Figure \textcolor{blue}{B2} occurs in sync with the behavior changes in planet 5 of simulation 157 discussed above. This shows how planet position is important, but the exchange of orbital angular momentum between planets does not always immediately cause changes in spin.

The other example of a very long initial regime in T.L. Zero ending with sporadic regime behaviors is seen in simulation number 132, planet 4, with Q=100 (see Table \textcolor{blue}{D3}). The spin integration begins with an 80,000 year long T.L. Zero regime. This example is even more interesting given that with Q=10, the planet was in the T.L. Zero regime for the entire integration. This is one of only two examples where the increase from Q=10 to Q=100 caused the TLSR spin state when Q=10 had only tidally locked regimes. Simulation number 182 of planet 4 is the other example where increasing to Q=100 caused the TLSR spin state. Although simulation 182 showed an initial T.L. Zero regime of about 47,000 years (Table \textcolor{blue}{D3}), the initial T.L. Zero regime in this simulation is not the longest of all its regimes and, thus, is a bit less impressive.

The last spin history we will explore is that of simulation number 132, planet 2. Information on planet 2 is not included in any tables or figures for brevity. Planet 2 is also excluded, in part, because TRAPPIST-1c may be the least habitable of the four planets we examine (see Section \ref{sec:EBMresults}). However, we should mention that the same three N-body simulations, of the total ten, that exhibit the TLSR spin state for planet 3 also exhibit the TLSR spin state for planet 2, both when Q=10 and Q=100. Simulation 132, with Q=10, is the only example where the T.L. Zero regime does not occur at all. This planet immediately begins spinning, similar to planet 5 in simulation 157. However, for Q=10, this initial TLSR-like behavior lasts for about 70 yrs before the planet enters a T.L. Pi regime for the remainder of integration. About 25 yrs of the initial 70 yrs are in the PTB regime which ends with the final T.L. Pi regime. For Q=100, the initial TLSR-like behavior lasts about 1000 yrs before the planet enters a T.L. Pi regime for the remaining integration time. For Q=100, all regime behaviors are seen in the initial 1000 yrs. These two simulations began their spin integrations in a period of high orbital forcing on their spins. Thus, it is unclear if these planets are actually in the TLSR spin state. These two simulations may be examples of boundary cases between the TLSR spin state and the tidally locked state.

\subsection{Energy Balance Model}
\label{sec:EBMresults}

For illustrative purposes, we retrieve the stellar insolation of each planet using the recommended stellar insolation of TRAPPIST-1e of $900\text{ } W/m^2$ from \citet{Fauchez2020} and the semi-major axes from \citet{Gillon2017}. The stellar insolations of planets c, d, e, and f are (in $W/m^2$) about $3000$, $1500$, $900$, and $500$, respectively. For reference, the solar insolations of Venus, Earth, and Mars are (in $W/m^2$) roughly $2600$, $1360$, and $590$, respectively. Of the four planets we examine, TRAPPIST-1c is the most likely to have entered into a runaway greenhouse at some point in the past due to its high stellar insolation (see Figure \textcolor{blue}{C1}). 

The temperature profile of these planets has rotated $180^{\circ}$, so many parts of the surface have experienced changes in temperature of 50K or more (see Figures \textcolor{blue}{C1}-\textcolor{blue}{C4}). The smallest change in temperature over the surface would be at the terminators (longitudes of $\pm90^{\circ}$) which are largely unchanged after rotation. However, the terminators would still experience large temperature changes during the planet's rotation, except at the planet's poles. The temperature change experienced during the course of the rotation is important to keep in mind as our figures only capture the final equilibrium state. The temperature change that each longitude experiences during rotation is extreme compared to anything seen on Earth.

In our EBMs, the top of atmosphere (TOA) albedo partly counteracts the effect of the constant surface albedo due to atmospheric concentrations of greenhouse gases at different temperatures. In addition, TRAPPIST-1's spectrum peaks stronger in the infrared and this causes a lower albedo of ice than on the Earth. As expected, these two factors make the difference in substellar temperature, $\Delta T_{sub}$, between $180^{\circ}$ rotations relatively small. The largest $\Delta T_{sub}$ is seen on TRAPPIST-1e (Figure \textcolor{blue}{C3}) with a change of about 2.99K. TRAPPIST-1d (Figure \textcolor{blue}{C2}) has the second largest $\Delta T_{sub}$ of 2.52K. TRAPPIST-1d and TRAPPIST-1e are the two most likely habitable planets in terms of stellar insolation compared to the Earth.

The relatively small $\Delta T_{sub}$s from our EBMs show that increases in albedo from ice, due to the TLSR spin state, is unlikely to force the TRAPPIST-1 planets into a snowball climate state. If the substellar temperature remains above freezing despite increased albedo, the day side ice sheets would eventually melt after a transition from T.L. Pi to T.L. Zero, or vice versa. Transitions from a tidally locked regime to a Spin regime would require models that account for active ice sheet formation and melting. Our EBMs have relatively small $\Delta T_{sub}$s in the sense that the temperature change is so small that it is only able to lower temperatures below the freezing point of sea ice in boundary cases where the initial substellar temperature is already a few Kelvin from the freezing temperature. Boundary cases have $\Delta T_{sub}$ which may be small enough that more complex climate systems could push the outcome either way. In other words, the $\Delta T_{sub}$ in our models is too small for a basic climate model to predict whether flipping a tidally locked planet, that lies around the boundary case, may send it into a snowball state. The boundary case is nearly present for our model of TRAPPIST-1f (see Figure \textcolor{blue}{C4}) where the substellar temperatures are close to the freezing point of sea ice. More robust climate models may be able to distinguish whether small decreases in substellar temperature at these boundary cases are actually able to cause a snowball state or if other factors are more important.

Although rather simple, our EBMs achieve the three goals outlined in section \ref{EBM}. First, we obtain a quantitative measure of the temperature change at each longitude in Figures \textcolor{blue}{C1}-\textcolor{blue}{C4}. Second, we determine the $\Delta T_{sub}$ due to the increased albedo of ice sheets is relatively small for M-dwarfs. Third, the increase in albedo from the TLSR state is unlikely to force the planet into a snowball state. Our EBM results demonstrate that the TLSR spin state is most impactful on habitability because of the temperature change that each longitude experiences over the course of a spin period.

\section{Discussion}
\label{sec:discussion}

Planets that exhibit TLSR spin states would experience extreme climate change when they transition between spin regimes. Transitions between two tidally locked regimes, with a final change in orientation of $180^{\circ}$, will flip the surface temperature profile of the planet. Transitions between a tidally locked regime and a Spinning regime will begin, or end, a day/ night cycle with a stellar day lasting on the order of an Earth year. In this section, we will discuss the larger implications of our results on climate processes that EBMs cannot model, geology, and habitability. We make inferences on these implications by comparing with similar changes in climate on Earth and the resulting phenomenon observed in Earth's recent and distant history.

The effect of substellar temperature change, $\Delta T_{sub}$, on climate is difficult to infer since previous climate models of tidally locked planets show that the substellar region is often covered in clouds from increased evaporation \citep{Lewis2018} and, likewise, powers hurricanes \citep{Yan2020}. In our model, a previously tidally locked planet that has rotated $180^{\circ}$ will have remnant ice sheets covering the substellar point. Lower substellar temperatures and the lack of an open ocean will reduce evaporation on the day side. Precipitation rates and hurricanes would presumably be reduced as a consequence. Thus, reduced evaporation on the day side, due to a substellar ice sheet, may reduce the transport of water from the day side to the night side of a planet that is tidally locked. Reduced atmospheric water transport would be impactful in reducing precipitation and, thus, reducing the speed that large continental ice sheets form on the night side, such as examined by \citet{Menou2013} and \citet{Yang2014}. Therefore, the night side may be spared a large buildup of ice sheets, at least until a substellar ocean reemerges and evaporation increases to normal levels. If the planet flips again, or begins spinning, soon after the previous flip then the negative habitability impacts of rapid ice sheet formation may be partly avoided. Negative impacts of rapid ice sheet formation include, but are not limited to, many meters of snowpack burying ecosystems.

Ocean currents and prevailing winds would change, given enough time, as these are ultimately driven by heat transport. On Earth, changes in ocean currents occur due to continental drift which is a much longer timescale than TLSR planets. One analogy we routinely observe on Earth are the El Niño and La Niña cycles in the Pacific Ocean or the change in weather patterns due to seasons. However, neither of these is nearly as extreme as what our models describe and they do not persist for years to millennia or longer. Earth's Milankovitch cycles, which are responsible for cyclic ice ages, are more similar in terms of being caused by orbital dynamics and the magnitude of temperature change. However, the timescale of Earth's Milankovitch cycles is around 50,000-100,000 years. Furthermore, local or regional changes in climate that coincide with different geographies are another issue entirely. In short, there is no equivalent analog we can observe on Earth and 3D GCMs are the best tool we have to study this phenomenon.

Extreme temperature changes affect geology through increased erosion. The most basic form of erosion would be stress from thermal expansion and the cracking of rock through ice intrusion. However, the potential rapid formation and retreat of glaciers would cause erosion directly but also through indirect phenomena like megafloods. Megafloods are caused when large glacial melt lakes rapidly discharge when ice dams collapse. They are likely responsible for many unique surface features in the northwest of the United States \citep{OCONNOR2020}. Another erosion source would be permafrost melt. Permafrost melt due to Earth's current human-caused climate change is observed to create gas blowouts \citep{Bogo2020}, large sinkholes, and landslides \citep{Fortier2007}. Overall, these erosive processes may degrade mountains more quickly than on Earth and cause accordingly flatter surface topologies. 

The consequences of TLSR spin states on habitability are largely apparent. The surface of these planets would experience extreme changes in weather and temperature during rotation which may make it difficult for any complex surface life to survive if it was present. Photosynthetic life in particular would need some mechanism to proliferate despite the loss of starlight. This may be through some form of dormancy or simply outpacing the rotation of the planet through dispersing seeds or spores, for example. Given the spin periods of our results, life may have a year or more to move and adjust. Comparing these planets to something like Earth, most of the consequences are negative. However, there may be some less obvious impacts that partly mitigate the negative. For example, increased erosion on land would mean an increase in nutrients in the oceans. Furthermore, high erosion may lead to quicker processing of rock into more suitable soil that allows faster resettling of life on recently deglaciated land. Reduced evaporation and precipitation on the day side due to ice sheets may also be partly counteracted by glacial runoff, at least locally and regionally. 

The high fraction of time spent in quasi-stable regime lengths, about 29\%, may also be a mitigating factor. Longer, or more prevalent, quasi-stable periods may allow diverse adaptations to take hold that reduce the impacts of the TLSR spin state to the level of an inconvenience. Although unlikely, the TLSR spin state could have unexpected benefits that outweigh the costs when compared to permanently tidally locked planets, much like Earth's seasons may be more beneficial than having stable temperatures throughout the year. Perhaps it is possible for life on TLSR planets to become accustomed to such frequent and extreme changes in temperature and climate. However, the temperature variations that a TLSR planet experiences are more extreme than any known period on Earth -- excluding the short periods that coincide with mass extinctions.

Our EBM results show that the TLSR spin state is unlikely to cause a permanent snowball state through increased albedo from ice. This differs from \citet{Chen2023}'s results showing that the TLSR spin state can cause a permanent snowball state in the outer habitable zone. Our EBMs differ from \citet{Chen2023}'s GCMs in model complexity and certain assumptions. The primary cause of our different results is our prescribed 0.25 albedo of water ice derived from TRAPPIST-1's spectrum by the THAI protocol \citep{Fauchez2020}. \citet{Chen2023}'s TOA on an ice-covered day side may be significantly higher than ours due to ExoPlaSim's gray cloud model \citep{Paradise2022}. \citet{Chen2023} reports high substellar albedo from clouds, with a TOA of 0.8-0.9, which our EBM does not have. Higher albedo on an ice-covered day side will lead to a larger $\Delta T_{sub}$ after a $180^{\circ}$ rotation and increase the chance of the planet entering a permanent snowball state. Determining the conditions where the TLSR spin state can cause a permanent snowball, and further study of climate impacts using a GCM, will be an objective of future work.

If any of the TRAPPIST-1 planets are indeed in a TLSR state, and if their atmospheric conditions are similar to what we assume in our EBMs, then albedo fluctuations may be observable. This would be contingent on if a planet still has a large albedo dichotomy from a previous tidally locked regime when it begins rotating. If the planet has a high fraction of land coverage, the similar albedo of land and ice around M-dwarfs (0.3 and 0.25, respectively) could cause confusion. The rarity of these observable albedo fluctuations would be subject to the longevity of night side ice sheets as rotation begins and the recent spin history of the planet. Other exoplanet systems may also be subject to TLSR spin states and our findings could also be applicable.

Lastly, it is important for one to keep in mind some of the spin histories shown in subsection \ref{sec:SpinResults} when discussing EBM results and implications. The discussion here has a wider focus on the initial changes experienced when a long-term persistent state comes to an end. The spin histories discussed in subsection \ref{sec:SpinResults} highlight some of the most acute applications of our discussion above. However, these implications are applicable to any spin integration with the TLSR spin state: to varying degrees. Table \ref{tab:avgReg}, Tables \textcolor{blue}{D6}-\textcolor{blue}{D1}, and a mean quasi-stable time fraction of 29\% show that regimes longer than 900 years are common and these regimes often end in a single libration or rotation period. It is worth mentioning that while many spin integrations do not exhibit the TLSR spin state over the time-span sampled, this may not be indicative of longer time-spans. The simulations highlighted in subsection \ref{sec:SpinResults}, especially simulations 132 and 182 for planet 4, exhibit how spin behavior depends on initial conditions and assumptions. Some simulations may exhibit different behaviors in preceding or subsequent spans of time, especially with different initial conditions. Thus, reliably predicting the exact current spin behavior, or recent history, in TRAPPIST-1 and similar systems would depend on determining the probability of the planets being in a specific behavior in best-fit simulations. Albedo fluctuations, or other indicators of rotation/ lack thereof, would make accurate predictions for specific planets easier.

\section{Conclusions}
\label{sec:conclusions}

Our results agree with previous findings of the TLSR spin state in systems similar to TRAPPIST-1. The types of spin histories and the average periods found here are similar to those found in \citet{Vinson2019} despite differences in the N-body simulations that we use. Our spin integrations show that spin history is dependent on the unique architecture of a given planetary system, the specific planet's position, and tidal parameters such as mass, radius, and Q-value. These findings help to achieve the objective of describing the complex, changing spin behaviors that TLSR planets undergo.

The regime length and rotation periods are defined and analyzed to classify the potential impact on climate. Our findings show that TLSR planets are able to be in both long-term persistent regimes and PTB regimes -– where frequent transitions between behaviors are present. Quasi-stable spin regimes, which last for 900 years or more, account for about 29\% of spin integration /textcolor{red}{time}, on average, when the TLSR spin state is observed. Long or quasi-stable regimes may be able to form persistent climate systems while the planet exhibits a single spin behavior for hundreds of years. On the high end, regimes that last for hundreds of millennia are almost certainly capable of persistent climate states. PTB and short-term spin regimes can cause interruptions in long-term regimes. These interruptions can lead to a unique scenario where a planet's rotation, with respect to its star, can reverse directions and change in nature sporadically. Interruptions in long-term regimes will lead to abrupt changes in stellar insolation, with respect to longitude, and may make consistent climate systems difficult to maintain through spin transitions. Climate systems that can be affected include ocean currents, prevailing winds, and precipitation patterns. Our analyses of rotation/libration periods show that transitions between spin behaviors occur on the scale of Earth years.

1D EBMs show that TLSR planets around M-dwarfs will experience a relatively small $\Delta T_{sub}$ due to the lower albedo of ice in an infrared dominant stellar spectrum. We find the largest $\Delta T_{sub}$ for TRAPPIST-1e at 2.99K which is only able to lower temperatures below freezing when a planet's substellar point is already near the freezing point. With such a small $\Delta T_{sub}$ from ice albedo, more complex climate systems may be more important in determining outcomes. Thus, substellar ice sheets, on their own, are unlikely to push a planet into a snowball state due to their small impact on $\Delta T_{sub}$. \citet{Chen2023}'s results show that the TLSR spin state can cause planets in the outer habitable zone to enter a permanent snowball state due to the planet's spin history. Our models do not form permanent snowballs primarily because of our prescribed water ice albedo of 0.25 for TRAPPIST-1 from the THAI protocol \citep{Fauchez2020}. Our models also differ due to ExoPlaSim's gray cloud model that can cause a high TOA albedo at the substellar point \citep{Paradise2022}. Future studies using a GCM will better determine when the TLSR spin state can cause a permanent snowball.

$\Delta T_{sub}$s on the order of a few Kelvin may have subtle impacts on evaporation and precipitation. However, if ice sheets from the previous night side are able to persist after a spin regime transition, they may be a stronger factor in affecting precipitation patterns. The more extreme change is in the temperature of different longitudes as the planet transitions from a tidally locked regime to a Spinning regime or after the planet flips $180^{\circ}$ and remains tidally locked. Rotating planets experience temperature changes at the equator of 50K or more over a single rotation period. The exact effects require more robust climate models, like 3D GCMs, to properly examine. However, using comparisons with climate changes on Earth, it is likely that erosion of land masses would increase and major climate systems would experience significant changes.

\section*{Acknowledgements}

The authors thank Nicolas Cowan for useful conversations on exoplanet climates. We also thank the anonymous reviewer for their feedback that improved our paper. Computer support was provided by UNLV’s National Supercomputing Center. CJS and JHS acknowledge support from the NSF grant AST-1910955. This material is based upon work supported by the National Aeronautics and Space Administration under the Nevada Space Grant Consortium Grant No. 13584.

\section*{Data Availability}

N-body simulations in this paper made use of the rebound code and reboundX which can be downloaded freely at http://github.com/hannorein/rebound and https://github.com/dtamayo/reboundx, respectively. The source code for the HEXTOR EBM can be downloaded freely at https://github.com/BlueMarbleSpace/hextor/releases. The N-body output data and the codes used for spin integration, post-processing, and visualization will be available upon publication at https://github.com/CodyShake/DayNite.



\bibliographystyle{mnras}
\bibliography{references} 





\appendix

\section{Derivation of Probability Equation}
\label{sec:appDerive}

There is a 100\% chance that a spin regime lands somewhere between the minimum regime length, $R$, and the simulation length, $S$,
\begin{equation}
    c_S \int_R^S f(x)dx = 1.
    \label{eq:App1}
\end{equation}
\noindent Here, f(x) is the power law of the desired regime type of the form
\begin{equation}
    f(x)=ax^m
\end{equation}
\noindent and $c_S$ is a constant to ensure the sum of all possibilities is equal to $1$. If we want to find the probability of any regime greater than a length, $L$, occurring, we use
\begin{equation}
    c_S \int_L^S f(x)dx = P,
    \label{eq:App2}
\end{equation}
\noindent where $P$ is the probability. Here, we assume that the cumulative distribution function of all our sample simulations, which exhibit the TLSR spin state, is roughly representative of all planets in the TLSR spin state. Furthermore, we assume this is representative only of TLSR planets near the habitable zone of stars that have a similar mass to TRAPPIST-1.

In the case discussed in section \ref{sec:SpinResults}, we choose the probability of the desired regime length beforehand. This is because we want to know the simulation length, $S$, required to produce regimes longer than a predefined $L$ at the same rate as our simulations produce regimes longer than 100,000 yrs. However, as long as $R<L<S$, one can solve the following equations for other quantities if desired.

We find the probability of regimes longer than 100 kyr at 12.8\% for simulations with a total length of 200 kyr. We find the probability by first evaluating Equation \ref{eq:App1} to find the value of $c_S$ when $R=10 yrs$ and $S=200 kyr$. We then evaluate Equation \ref{eq:App2} to find the probability $P$ when $L>100 kyr$ (i.e. $P(L>100 kyr)=0.128$).

Above, we solve for the probability since we know the total length of the simulations we run. Now, we want to find the total simulation length needed for a 1 million-year regime to occur at the same probability. Since $c_S$ is the same in the two equations, we can combine the two to retrieve a solvable equation
\begin{equation}
    \frac{1}{\int_R^S f(x)dx} = \frac{P}{\int_L^S f(x)dx}.
\end{equation}
\noindent As long as the power, $m$, of the power law is not exactly equal to $1$, then the integral's result has the general form
\begin{equation}
    \int_a^b ax^m dx = \frac{a}{m+1}(b^{m+1}-a^{m+1}).
    \label{eq:genInt}
\end{equation}
\noindent After evaluating each integral and solving for $S$, we retrieve the final equation that is shown in section \ref{sec:SpinResults},
\begin{equation}
    S=\left( \frac{L^{m+1}-PR^{m+1}}{1-P} \right)^{\frac{1}{m+1}}.
\end{equation}
\noindent The coefficient of $\frac{a}{m+1}$ in Eq. \ref{eq:genInt} cancels out in the algebra which means the scaling factor of the power law, $a$, does not impact the result. This is consistent as the scaling factor should be irrelevant here as the value of $a$ depends on the total number of regimes, which correlates to the simulation length, and the bin width of the histogram, which is subjective. Indeed, the value of $m$ remains constant, while $a$ varies, when the histogram and fit are conducted with different bin widths.

\clearpage

\includepdf[pages=-]{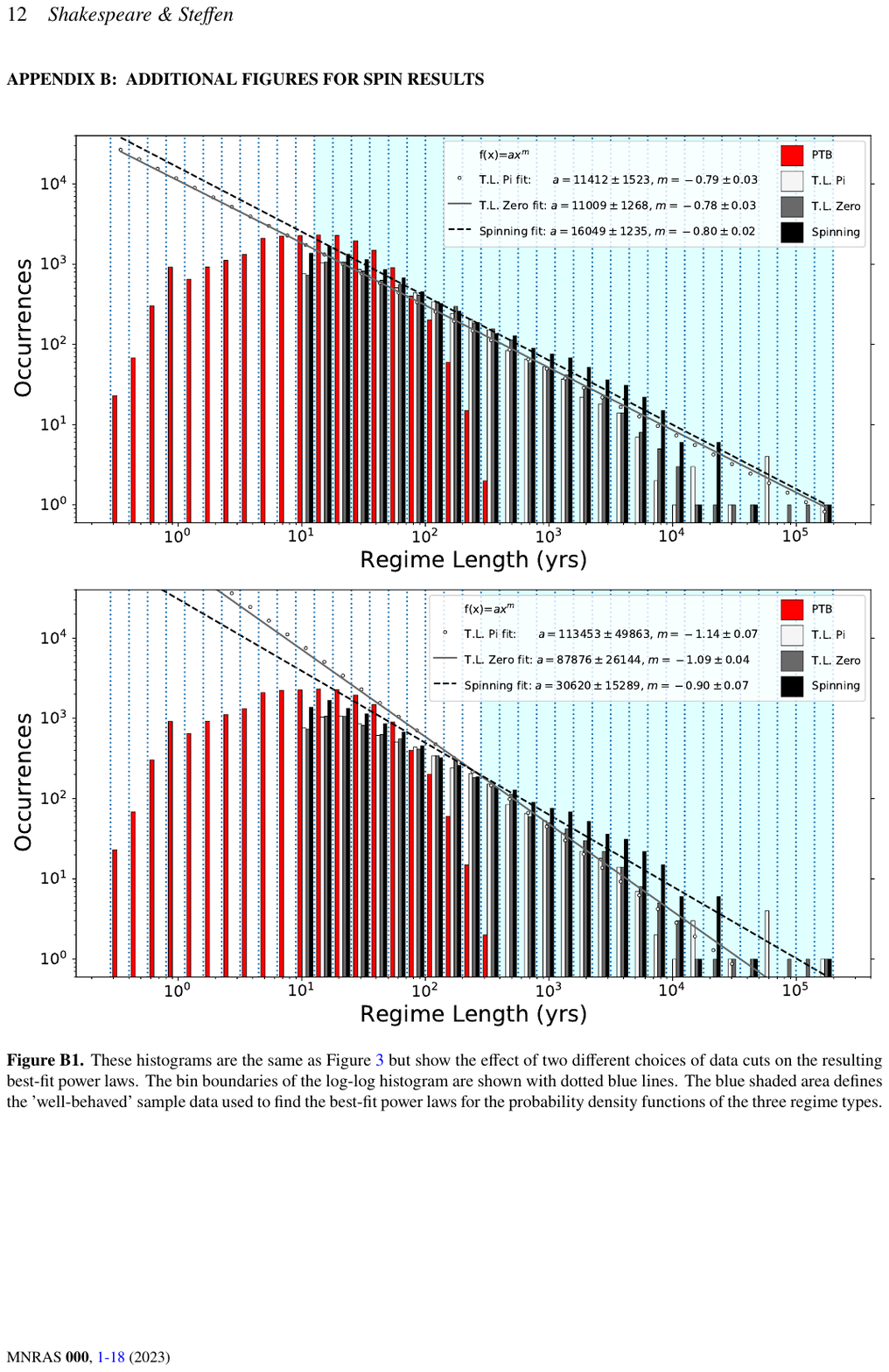}

\bsp	
\label{lastpage}
\end{document}


\pagestyle{asd}

%
\section{ADDITIONAL FIGURES FOR SPIN RESULTS}
\label{sec:appSpin}

\begin{figure}[H]
    \centering
    \includegraphics[width=\textwidth]{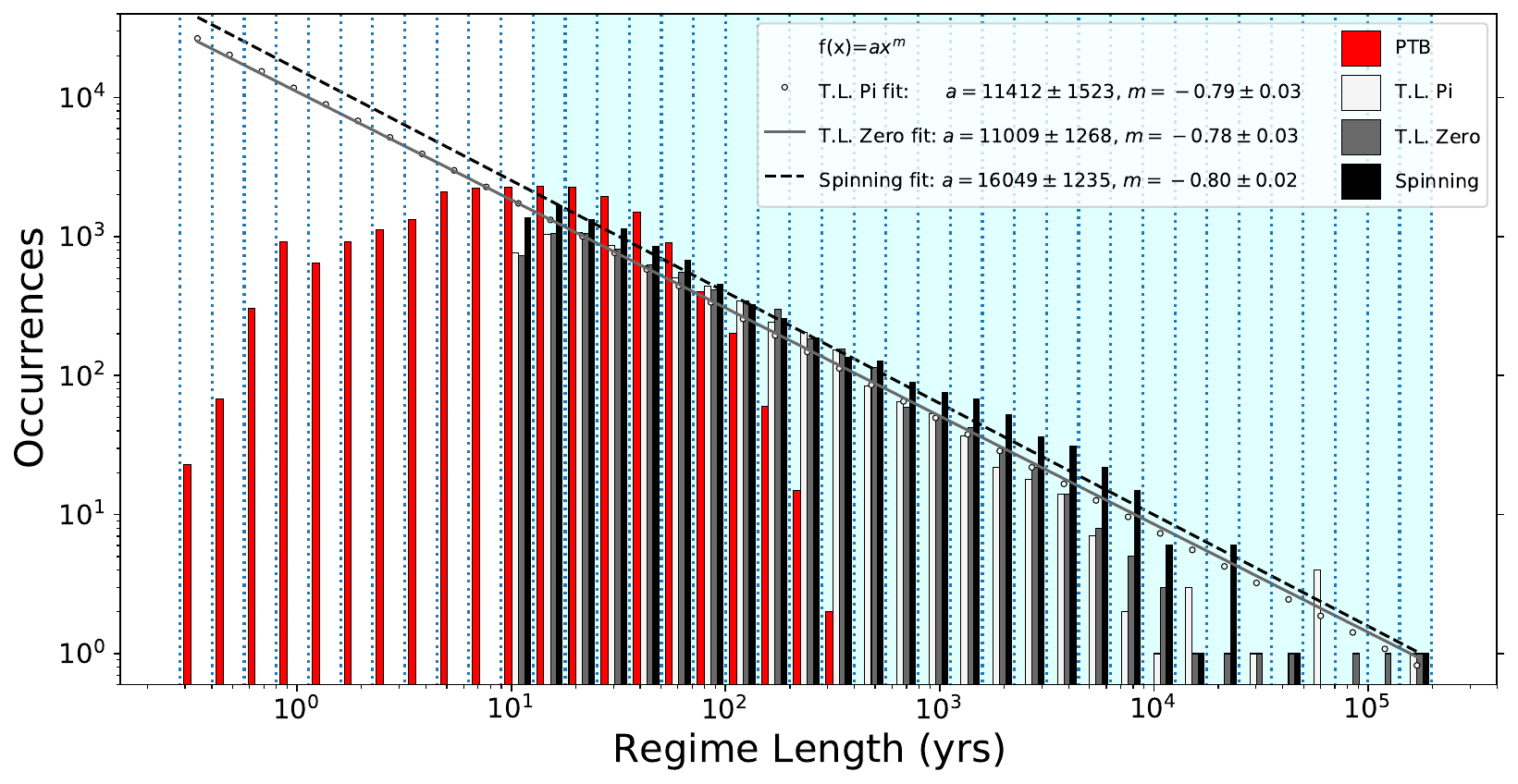}
    \includegraphics[width=\textwidth]{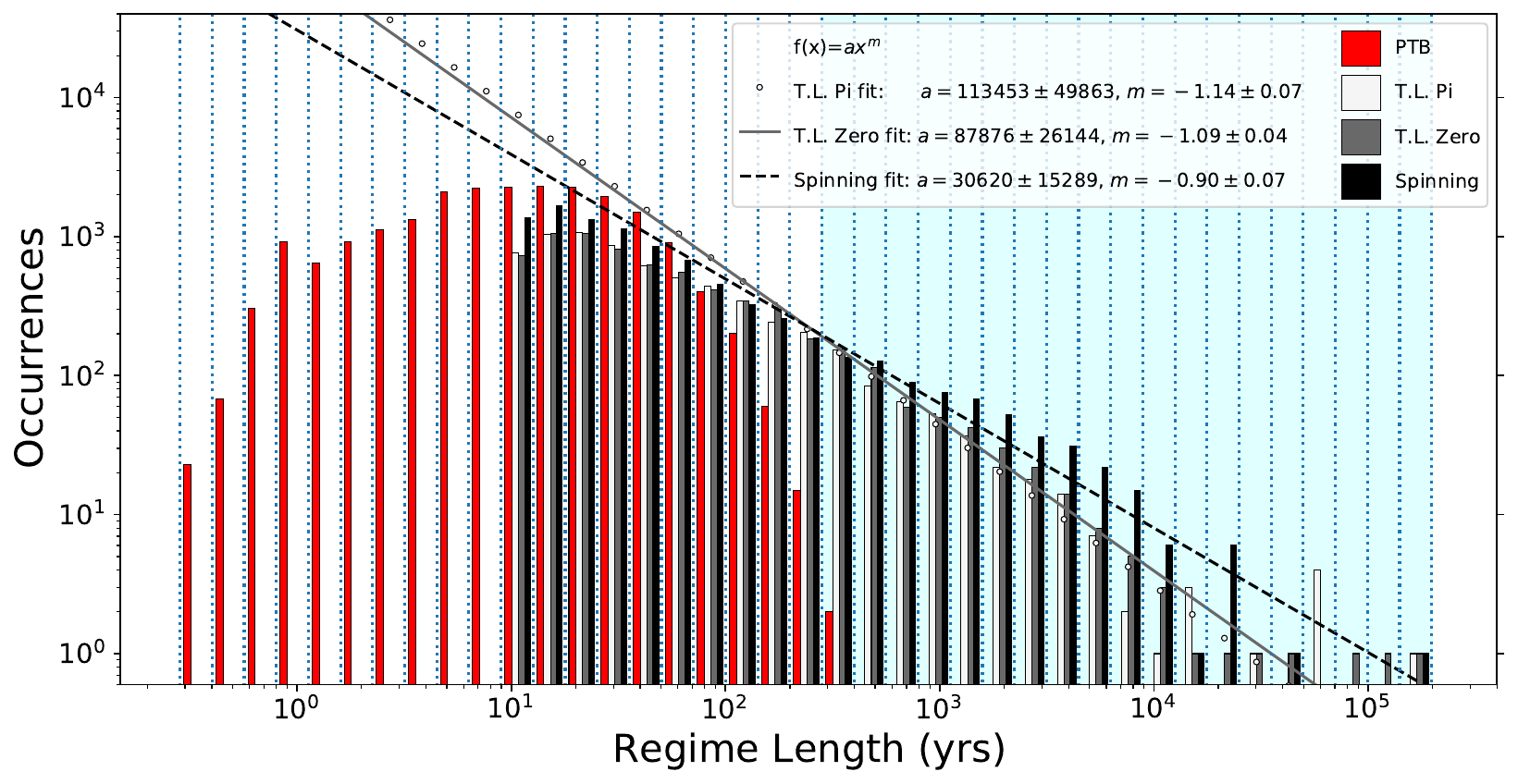}
    \caption{These histograms are the same as Figure \textcolor{blue}{3} but show the effect of two different choices of data cuts on the resulting best-fit power laws. The bin boundaries of the log-log histogram are shown with dotted blue lines. The blue shaded area defines the 'well-behaved' sample data used to find the best-fit power laws for the probability density functions of the three regime types.}
    \label{fig:HistMinMax}
    \vspace{2in}
\end{figure}

\begin{figure}[H]
    \centering
    \includegraphics[width=\textwidth]{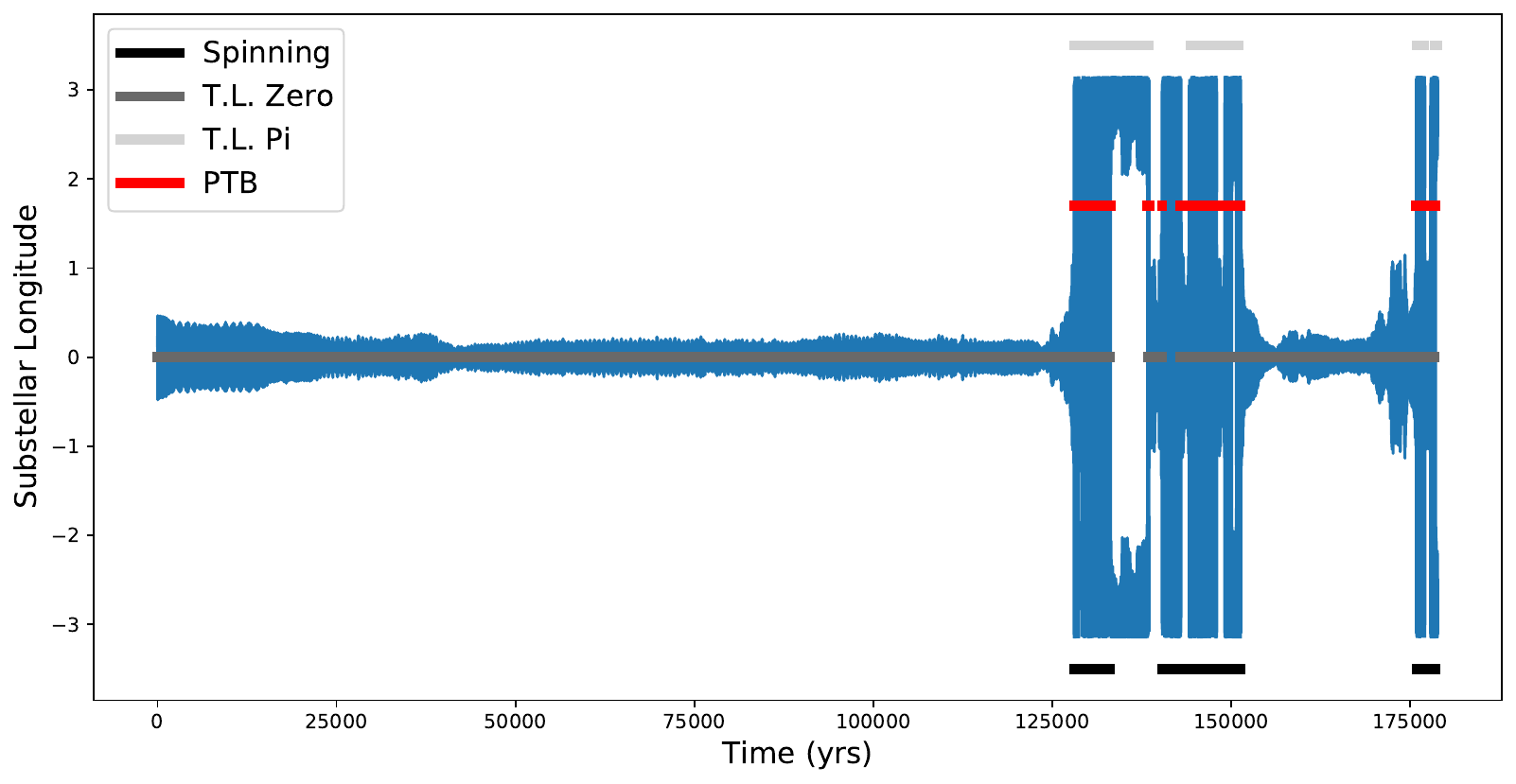}
    \includegraphics[width=\textwidth]{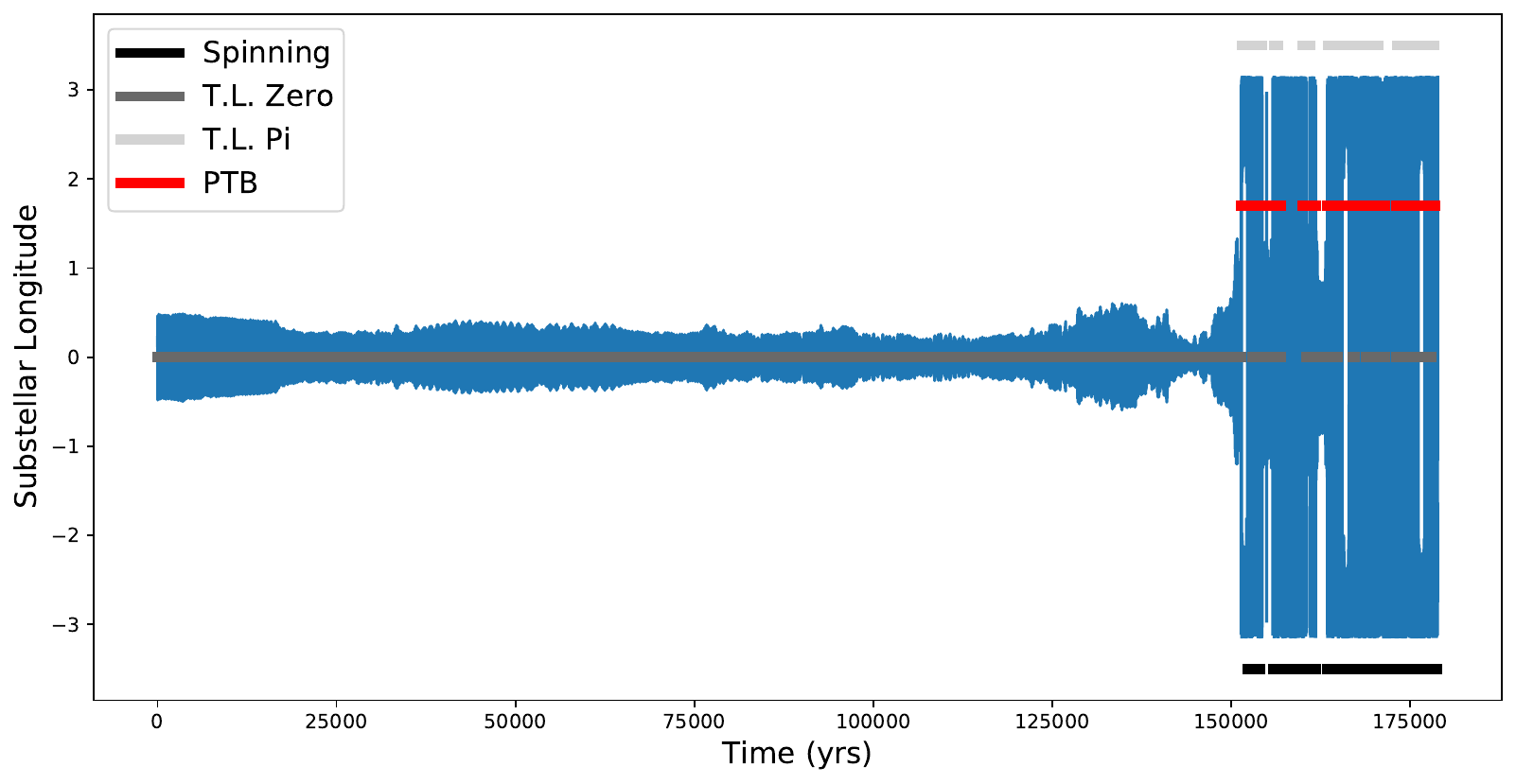}
    \caption{Spin history and regime classification of simulation number 157 of planet 4 with (Top) Q=10 and (Bottom) Q=100. Blue lines represent the evolution of the substellar longitude which can wrap from $-\pi$ to $\pi$. Solid, vertical blue lines are indicative of the Spinning or PTB regimes. Colored horizontal lines at $-\pi$, zero, two, and $\pi$ show regime classification.}
    \label{fig:spin157p4}
\end{figure}

\pagebreak

%
\section{FIGURES FOR EBM RESULTS}
\label{sec:appEBM}
\setcounter{figure}{0} 

\begin{figure}[H]
\begin{minipage}{0.5\linewidth}
    \centering
    \includegraphics[width=0.8\columnwidth]{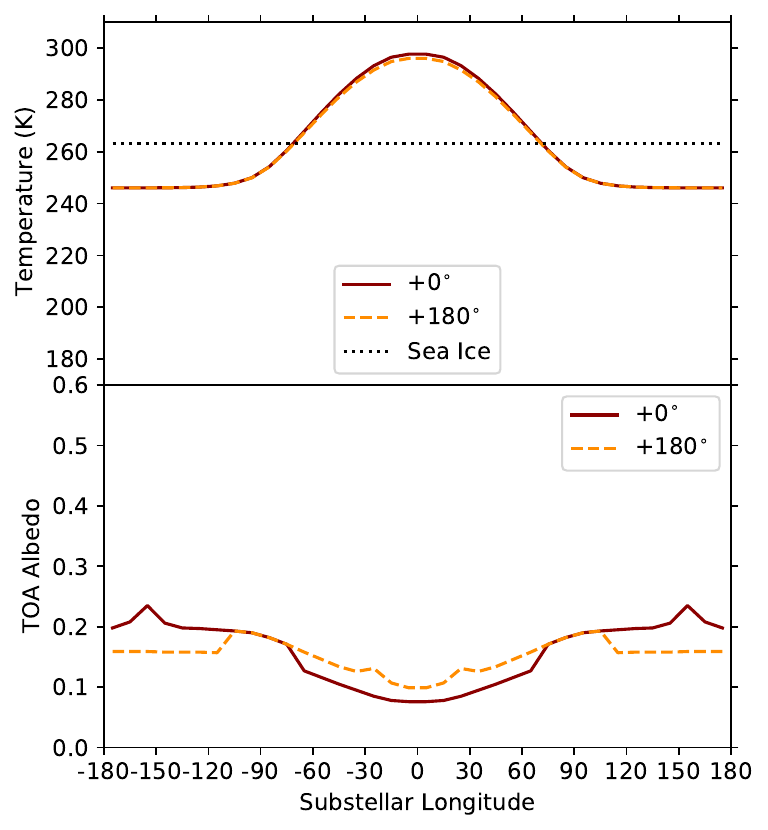}
    \caption{Temperature and TOA albedo profile of TRAPPIST-1c with respect to substellar longitude. Ice has an albedo 0.25 and sea ice freezes at 263.15K (dotted line). This planet experiences a 1.63K change in substellar temperature.}
    \label{fig:TempC}
\end{minipage}
\hspace{0.25in}
\begin{minipage}{0.5\linewidth}
    \centering
    \includegraphics[width=0.8\columnwidth]{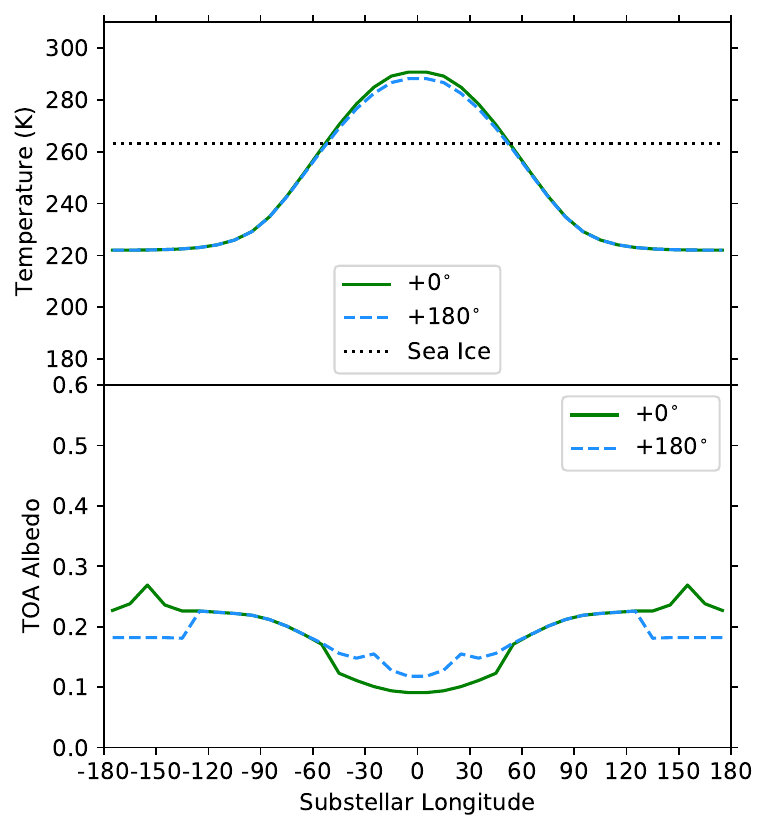}
    \caption{Temperature and TOA albedo profile of TRAPPIST-1d with respect to substellar longitude. Ice has an albedo 0.25 and sea ice freezes at 263.15K (dotted line). This planet experiences a 2.52K change in substellar temperature.}
    \label{fig:TempD}
\end{minipage}
\end{figure}

\begin{figure}[H]
\begin{minipage}{0.5\linewidth}
    \centering
    \includegraphics[width=0.8\columnwidth]{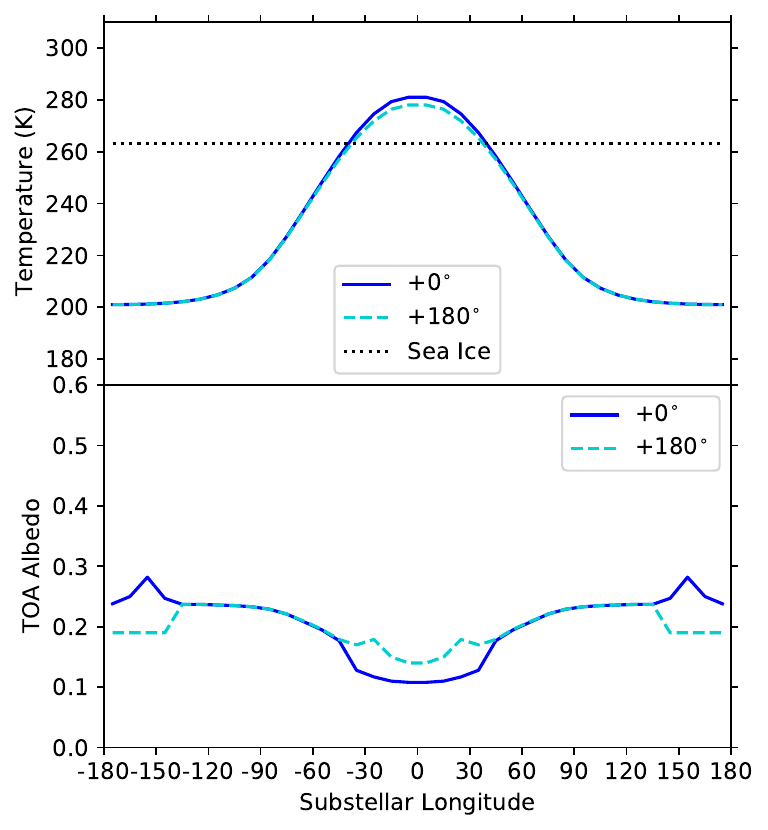}
    \caption{Temperature and TOA albedo profile of TRAPPIST-1e with respect to substellar longitude. Ice has an albedo 0.25 and sea ice freezes at 263.15K (dotted line). This planet experiences a 2.99K change in substellar temperature.}
    \label{fig:TempE}
\end{minipage}
\hspace{0.25in}
\begin{minipage}{0.5\linewidth}
    \centering
    \includegraphics[width=0.8\columnwidth]{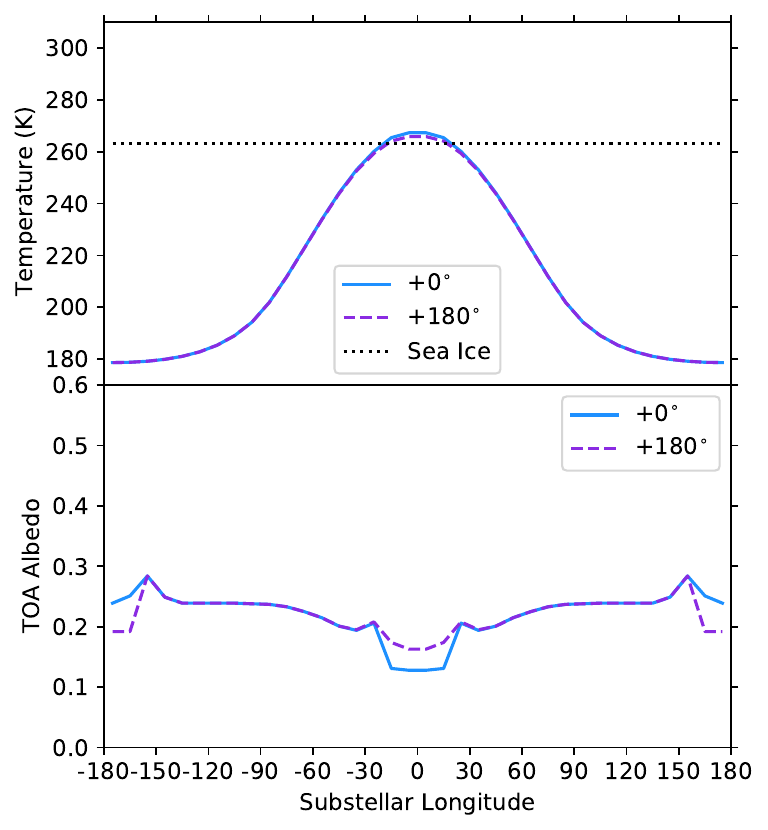}
    \caption{Temperature and TOA albedo profile of TRAPPIST-1f with respect to substellar longitude. Ice has an albedo 0.25 and sea ice freezes at 263.15K (dotted line). This planet experiences a 1.48K change in substellar temperature.}
    \label{fig:TempF}
    \end{minipage}
\end{figure}

%

\clearpage
\onecolumn
\section{DATA TABLES FOR SPIN RESULTS}
\label{sec:appTAB}
\captionsetup{skip=-5pt, singlelinecheck=false}

\renewcommand{\figurename}{\textbf{Table \thesection\thetable.}}
\renewcommand{\arraystretch}{1.2}

\begin{table*}[!htbp]
    \scriptsize
	\centering
	\caption{Regime statistics of Planet 5. Simulations 17, 44, 132, 139, 182, 245, and 252 are excluded as they remain in the 'T.L. Zero' regime for the entire duration of the simulation. All values are in years.}
	\label{tab:planet5reg}
    \begin{tabular}{lrrrrrrrr}
        \\\\
        \hline
        \multicolumn{2}{c}{Planet 5}&\multicolumn{3}{c}{Q=10}&\vline&\multicolumn{3}{c}{Q=100}\\
        \hline
        &Simulation&152&157&169&\vline&152&157&169\\
        \hline
        \hline
        \multirow{7}{*}{Spin}&Mean Len.&249.3&513.5&235.1&\vline&661.7&	2468.4&	357.5\\
        &Std. Dev.&705.7&2737.4&603.4&\vline&3357.4&	20702.7&	1296.1\\
        &Med. Len.&36.5&40.9&46.6&\vline&41.1&	28.9&	40.8\\
        &1st Quartile&20.0&22.6&23.8&\vline&21.8&	21.0&	22.0\\
        &3rd Quartile&113.1&89.4&109.2&\vline&141.1&	65.4&	99.7\\
        &Min. Len.&10.1&17.9&13.6&\vline&10.4&	17.7&	13.9\\
        &Max. Len.&6940.4&24424.7&5026.1&\vline&46775.3&	198046.9&	14140.0\\
        \hline
        \multirow{7}{*}{T.L. Zero}&Mean Len.&153.1&410.6&236.8&\vline&106.2&	206.4&	228.4\\
        &Std. Dev.&809.1&	1435.6&	575.8&\vline&503.6&	533.4&	481.4\\
        &Med. Len.&31.8&	22.7&	51.4&\vline&31.0&	34.2&	49.0\\
        &1st Quartile&18.8&	16.8&	22.4&\vline&18.7&	17.2&	22.9\\
        &3rd Quartile&70.1&	96.7&	164.2&\vline&63.9&	106.5&	143.1\\
        &Min. Len.&10.1&	10.1&	10.1&\vline&10.1&	10.1&	10.1\\
        &Max. Len.&11917.5&	10820.9&	5309.2&\vline&6204.0&	3904.6&	2907.2\\
        \hline
        \multirow{7}{*}{T.L. Pi}&Mean Len.&336.5&1787.7&206.8&\vline&125.0&	404.0&	169.3\\
        &Std. Dev.&3149.3&16635.9&481.1&\vline&559.3&	1287.7&	377.4\\
        &Med. Len.&26.7&27.9&43.5&\vline&29.6&	25.8&	44.0\\
        &1st Quartile&18.4&16.8&20.1&\vline&18.4&	18.2&	22.5\\
        &3rd Quartile&61.6&198.6&131.0&\vline&70.4&	260.3&	123.2\\
        &Min. Len.&10.1&10.1&10.1&\vline&10.1&	10.4&	10.1\\
        &Max. Len.&51710.7&175421.4&3975.6&\vline&5928.3&	10457.3&	4065.4\\
        \hline
        \multirow{7}{*}{PTB}&Mean Len.&20.9&22.8&15.0&\vline&21.6&	22.8&	13.8\\
        &Std. Dev.&21.8&	22.9&	14.3&\vline&20.6&	24.2&	13.1\\
        &Med. Len.&13.7&	14.6&	9.6&\vline&14.5&	15.7&	9.1\\
        &1st Quartile&6.9&	8.7&	5.9&\vline&6.9&	7.6&	5.1\\
        &3rd Quartile&28.8&	28.7&	21.1&\vline&29.9&	30.8&	18.7\\
        &Min. Len.&1.9&	1.4&	1.3&\vline&1.4&	1.7&	1.6\\
        &Max. Len.&181.8&	140.1&	100.2&\vline&128.6&	226.6&	83.4\\
        \hline
    \end{tabular}
\end{table*}

\begin{table*}[!htbp]
    \scriptsize
	\centering
	\caption{Period statistics of Planet 5. Q=100 simulations are excluded for brevity. For simulations that solely remain in the 'T.L. Zero' regime, the mean and median periods for Q=100 simulations are all greater than or equal to the Q=10 simulations. All values are in years.}
	\label{tab:planet5per}
    \begin{tabular}{lrrrrrrrrrrr}
        \\\\
        \hline
        \multicolumn{2}{c}{Planet 5}&\multicolumn{10}{c}{Q=10}\\
        \hline
        &Simulation&17&44&132&139&152&157&169&182&245&252\\
        \hline
        \hline
        \multirow{7}{*}{Spin}&Mean Per.&---	&---	&---	&---&	5.04&	4.28&	4.56	&---	&---	&---\\
        &Std. Dev.&---	&---	&---	&---&	1.33&	2.33&	1.50	&---	&---	&---\\
        &Med. Per.&---	&---	&---	&---&	4.94&	3.36&	4.27	&---	&---	&---\\
        &1st Quartile&---	&---	&---	&---&	3.84&	2.24&	3.73	&---	&---	&---\\
        &3rd Quartile&---	&---	&---	&---&	5.49&	5.60&	5.33	&---	&---	&---\\
        &Min. Per.&---	&---	&---	&---&	2.74&	1.68&	2.13	&---	&---	&---\\
        &Max. Per.&---	&---	&---	&---&	12.62&	12.89&	13.33	&---	&---	&---\\
        \hline
        \multirow{7}{*}{T.L. Zero}&Mean Per.&2.45	&2.52	&2.05	&1.12	&6.41&	6.85&	5.86&	2.27&	2.67&	5.09\\
        &Std. Dev.&1.72	&1.80	&1.65	&0.69	&1.18&	1.52&	1.32&	1.31&	1.75&	0.24\\
        &Med. Per.&1.67	&2.04	&1.10	&1.16	&6.03&	6.72&	5.86&	2.25&	2.20&	4.97\\
        &1st Quartile&1.11	&0.51	&0.55	&0.58	&5.49&	5.60&	5.33&	1.69&	1.10&	4.97\\
        &3rd Quartile&3.90	&4.60	&3.29	&1.16	&7.13&	7.84&	6.40&	2.25&	4.95&	4.97\\
        &Min. Per.&0.56	&0.51	&0.55	&0.58	&0.55&	5.04&	0.53&	0.56&	0.55&	4.42\\
        &Max. Per.&8.91	&6.13	&7.69	&6.97	&16.46&	27.46&	17.06&	7.87&	7.70&	5.52\\
        \hline
        \multirow{7}{*}{T.L. Pi}&Mean Per.&---	&---	&---	&---	&5.47&	5.18&	5.99	&---	&---	&---\\
        &Std. Dev.&---	&---	&---	&---	&1.39&	1.32&	1.28	&---	&---	&---\\
        &Med. Per.&---	&---	&---	&---	&5.49&	5.04&	5.86	&---	&---	&---\\
        &1st Quartile&---	&---	&---	&---	&4.94&	5.04&	5.33	&---	&---	&---\\
        &3rd Quartile&---	&---	&---	&---	&6.03&	5.60&	6.40	&---	&---	&---\\
        &Min. Per.&---	&---	&---	&---	&0.55&	0.56&	0.53	&---	&---	&---\\
        &Max. Per.&---	&---	&---	&---	&17.01&	18.49&	18.13	&---	&---	&---\\
        \hline
    \end{tabular}
\end{table*}

\begin{table*}[!htbp]
    \scriptsize
	\centering
	\caption{Planet 4 is the only example in which increasing the Q-value to 100 has a significant effect on spin history. Simulations 132 and 182 remain in the 'T.L. Zero' regime when Q=10 (not shown), but experience behavior in all regimes when Q=100. Q=10 simulations are excluded for brevity. All values are in years.}
	\label{tab:planet4reg}
    \begin{tabular}{lrrrrrrr}
        \\\\
        \hline
        \multicolumn{2}{c}{Planet 4}&\multicolumn{6}{c}{Q=100}\\
        \hline
        &Simulation&17&132&152&157&169&182\\
        \hline
        \hline
        \multirow{7}{*}{Spin}&Mean Len.&203.0&	38.9&	176.2&	108.9&	332.8&	54.0\\
        &Std. Dev.&630.6&	111.3&	530.1&	347.8&	1532.4&	214.2\\
        &Med. Len.&24.4&	19.6&	31.8&	31.7&	27.2&	12.5\\
        &1st Quartile&14.1&	14.2&	17.1&	13.1&	15.0&	11.4\\
        &3rd Quartile&86.1&	29.2&	86.5&	61.5&	67.1&	21.6\\
        &Min. Len.&10.0&	10.1&	10.1&	10.4&	10.1&	10.6\\
        &Max. Len.&4751.1&	1779.6&	6411.1&	2911.7&	18083.5&2218.8\\
        \hline
        \multirow{7}{*}{T.L. Zero}&Mean Len.&180.4&	406.9&	137.1&	2925.7&	57.2&	363.0\\
        &Std. Dev.&811.4&	5248.3&	473.6&	20390.0&57.8&	3684.9\\
        &Med. Len.&32.6&	24.3&	23.9&	36.6&	37.4&	26.5\\
        &1st Quartile&19.7&	15.0&	14.4&	15.4&	20.4&	18.5\\
        &3rd Quartile&66.3&	52.4&	66.7&	110.4&	70.8&	57.7\\
        &Min. Len.&10.0&	10.1&	10.1&	11.1&	10.1&	11.5\\
        &Max. Len.&8371.5&	80494.4&4400.5&	151355.6&485.3&	46796.3\\
        \hline
        \multirow{7}{*}{T.L. Pi}&Mean Len.&108.9&	112.1&	109.2&	94.1&	68.9&	503.1\\
        &Std. Dev.&484.1&	896.6&	330.4&	194.2&	81.7&	4798.4\\
        &Med. Len.&30.2&	23.3&	23.5&	28.5&	39.2&	28.6\\
        &1st Quartile&19.8&	14.7&	14.4&	17.6&	17.9&	19.8\\
        &3rd Quartile&53.2&	43.3&	57.1&	64.7&	83.5&	68.4\\
        &Min. Len.&10.4&	10.1&	10.1&	10.6&	10.1&	10.1\\
        &Max. Len.&7827.5&	13400.9&2959.2&	943.0&	763.2&	60242.6\\
        \hline
        \multirow{7}{*}{PTB}&Mean Len.&26.8&	21.5&	17.7&	25.0&	11.0&	32.0\\
        &Std. Dev.&35.1&	25.5&	21.8&	29.5&	11.4&	37.8\\
        &Med. Len.&14.2&	12.2&	9.7&	17.0&	7.4&	18.9\\
        &1st Quartile&7.3&	4.5&	3.8&	4.6&	3.9&	5.5\\
        &3rd Quartile&32.2&	28.7&	24.2&	33.1&	14.5&	41.8\\
        &Min. Len.&0.9&	1.1&	0.5&	1.3&	0.7&	1.1\\
        &Max. Len.&293.9&	206.5&	201.3&	252.5&	129.4&	209.9\\
        \hline
    \end{tabular}
\end{table*}

\begin{table*}[!htbp]
    \scriptsize
	\centering
	\caption{Period statistics of Planet 4. Q=10 simulations are excluded for brevity. All values are in years.}
	\label{tab:planet4per}
    \begin{tabular}{lrrrrrrrrrrr}
        \\\\
        \hline
        \multicolumn{2}{c}{Planet 4}&\multicolumn{10}{c}{Q=100}\\
        \hline
        &Simulation&17&44&132&139&152&157&169&182&245&252\\
        \hline
        \hline
        \multirow{7}{*}{Spin}&Mean Per.&4.18	&---&	4.56	&---&	1.86&	4.28&	1.94&	4.89	&---	&---\\
        &Std. Dev.&0.75	&---&	0.96	&---&	0.94&	0.90&	0.83&	0.96	&---	&---\\
        &Med. Per.&4.01	&---&	4.31	&---&	1.44&	4.29&	1.77&	4.77	&---	&---\\
        &1st Quartile&3.64	&---&	3.95	&---&	1.08&	3.58&	1.41&	4.40	&---	&---\\
        &3rd Quartile&4.37	&---&	5.03	&---&	2.51&	4.65&	2.12&	5.13	&---	&---\\
        &Min. Per.&2.55	&---&	2.87	&---&	0.72&	2.50&	1.06&	3.30	&---	&---\\
        &Max. Per.&11.29	&---&	10.06	&---&	8.26&	10.02&	8.84&	12.10	&---	&---\\
        \hline
        \multirow{7}{*}{T.L. Zero}&Mean Per.&4.05&	3.01&	3.91&	3.31&	3.76&	3.29&	3.89&	3.76&	3.64&	2.38\\
        &Std. Dev.&0.73&	0.61&	0.64&	0.27&	1.34&	0.55&	0.83&	0.75&	0.62&	1.47\\
        &Med. Per.&4.01&	3.03&	3.95&	3.43&	3.95&	3.22&	3.89&	3.67&	3.62&	2.17\\
        &1st Quartile&3.64&	3.03&	3.59&	3.05&	3.23&	3.22&	3.54&	3.30&	3.62&	1.08\\
        &3rd Quartile&4.37&	3.36&	4.31&	3.43&	4.31&	3.58&	4.24&	4.03&	3.98&	3.98\\
        &Min. Per.&0.36&	0.34&	0.36&	0.38&	0.36&	0.36&	0.35&	0.37&	0.36&	0.36\\
        &Max. Per.&12.74&	6.05&	11.86&	4.19&	13.65&	8.94&	10.25&	13.56&	7.23&	6.87\\
        \hline
        \multirow{7}{*}{T.L. Pi}&Mean Per.&4.08	&---&	4.74	&---&	3.88&	4.25&	3.80&	3.98	&---	&---\\
        &Std. Dev.&0.75	&---&	0.60	&---&	1.34&	0.70&	0.88&	0.67	&---	&---\\
        &Med. Per.&4.01	&---&	4.67	&---&	3.95&	3.93&	3.89&	4.03	&---	&---\\
        &1st Quartile&3.64	&---&	4.31	&---&	3.59&	3.93&	3.54&	3.67	&---	&---\\
        &3rd Quartile&4.37	&---&	5.03	&---&	4.49&	4.65&	4.24&	4.40	&---	&---\\
        &Min. Per.&1.82	&---&	3.95	&---&	0.36&	3.22&	0.35&	0.37	&---	&---\\
        &Max. Per.&13.47	&---&	12.58	&---&	11.85&	9.66&	13.79&	11.36	&---	&---\\
        \hline
    \end{tabular}
\end{table*}

\begin{table*}[!htbp]
    \scriptsize
	\centering
	\caption{Regime statistics of Planet 3. Simulations 17, 44, 139, 157, 182, 245, and 252 are excluded as they remain in the 'T.L. Zero' regime for the entire duration of the simulation. All values are in years.}
	\label{tab:planet3reg}
    \begin{tabular}{lrrrrrrrr}
        \\\\
        \hline
        \multicolumn{2}{c}{Planet 3}&\multicolumn{3}{c}{Q=10}&\vline&\multicolumn{3}{c}{Q=100}\\
        \hline
        &Simulation&132&152&169&\vline&132&152&169\\
        \hline
        \hline
        \multirow{7}{*}{Spin}&Mean Len.&75.3&	100.1&	55.1&\vline&	144.2&	247.9&	243.7\\
        &Std. Dev.&168.2&	282.2&	135.0&\vline&	566.7&	844.6&	819.4\\
        &Med. Len.&35.6&	21.6&	22.8&\vline&	36.5&	27.2&	25.9\\
        &1st Quartile&19.9&	13.0&	13.6&\vline&	17.2&	14.2&	14.8\\
        &3rd Quartile&71.6&	58.2&	51.3&\vline&	82.9&	61.5&	55.5\\
        &Min. Len.&10.1&	10.0&	10.1&\vline&	10.1&	10.0&	10.1\\
        &Max. Len.&1670.4&	3734.8&	2568.0&\vline&	6088.5&	7693.5&	6698.4\\
        \hline
        \multirow{7}{*}{T.L. Zero}&Mean Len.&159.2&	65.5&	94.0&\vline&	130.1&	60.2&	72.1\\
        &Std. Dev.&1395.3&	93.2&	186.3&\vline&	499.2&	83.5&	129.3\\
        &Med. Len.&27.1&	26.8&	22.7&\vline&	29.0&	26.2&	21.2\\
        &1st Quartile&15.3&	15.3&	14.1&\vline&	16.5&	14.7&	13.4\\
        &3rd Quartile&75.5&	58.9&	76.2&\vline&	86.1&	65.1&	53.9\\
        &Min. Len.&10.1&	10.0&	10.1&\vline&	10.1&	10.0&	10.1\\
        &Max. Len.&26063.7&513.6&	1656.2&\vline&	7141.8&	515.9&	786.6\\
        \hline
        \multirow{7}{*}{T.L. Pi}&Mean Len.&88.8&	68.7&	93.8&\vline&	79.2&	66.5&	90.3\\
        &Std. Dev.&419.8&	96.7&	201.3&\vline&	228.8&	100.0&	178.9\\
        &Med. Len.&29.6&	27.6&	25.1&\vline&	26.1&	26.6&	21.6\\
        &1st Quartile&16.5&	15.7&	15.0&\vline&	15.5&	14.4&	14.3\\
        &3rd Quartile&70.2&	72.0&	56.9&\vline&	66.8&	74.8&	79.4\\
        &Min. Len.&10.1&	10.0&	10.1&\vline&	10.1&	10.0&	10.1\\
        &Max. Len.&7718.7&	609.5&	1745.9&\vline&	3079.3&	643.7&	1349.1\\
        \hline
        \multirow{7}{*}{PTB}&Mean Len.&15.8&	13.7&	20.3&\vline&	14.7&	13.9&	21.3\\
        &Std. Dev.&18.3&	15.3&	21.2&\vline&	14.8&	15.0&	21.6\\
        &Med. Len.&9.4&	8.3&	13.8&\vline&	9.7&	9.2&	14.2\\
        &1st Quartile&4.9&	3.0&	5.3&\vline&	4.2&	3.2&	5.0\\
        &3rd Quartile&20.7&	19.1&	28.1&\vline&	20.6&	19.0&	31.4\\
        &Min. Len.&0.5&	0.6&	0.5&\vline&	0.6&	0.6&	0.5\\
        &Max. Len.&149.0&	109.9&	164.6&\vline&	97.0&	109.2&	116.8\\
        \hline
    \end{tabular}
\end{table*}

\begin{table*}[!htbp]
    \scriptsize
	\centering
	\caption{Period statistics of Planet 3. Q=100 simulations are excluded for brevity. All values are in years.}
	\label{tab:planet3per}
    \begin{tabular}{lrrrrrrrrrrr}
        \\\\
        \hline
        \multicolumn{2}{c}{Planet 3}&\multicolumn{10}{c}{Q=10}\\
        \hline
        &Simulation&17&44&132&139&152&157&169&182&245&252\\
        \hline
        \hline
        \multirow{7}{*}{Spin}&Mean Per.&---	&---&	2.71	&---&	2.04	&---&	2.33	&---	&---	&---\\
        &Std. Dev.&---	&---&	0.60	&---&	0.63	&---&	0.65	&---	&---	&---\\
        &Med. Per.&---	&---&	2.61	&---&	1.91	&---&	2.34	&---	&---	&---\\
        &1st Quartile&---	&---&	2.13	&---&	1.67	&---&	1.87	&---	&---	&---\\
        &3rd Quartile&---	&---&	3.08	&---&	2.38	&---&	2.57	&---	&---	&---\\
        &Min. Per.&---	&---&	1.66	&---&	0.95	&---&	0.94	&---	&---	&---\\
        &Max. Per.&---	&---&	5.92	&---&	7.63	&---&	6.31	&---	&---	&---\\
        \hline
        \multirow{7}{*}{T.L. Zero}&Mean Per.&1.85&	0.91&	2.27&	2.13&	2.63&	1.34&	2.49&	1.84&	1.62&	1.30\\
        &Std. Dev.&1.25&	0.82&	0.65&	0.56&	0.51&	0.59&	0.49&	0.79&	0.84&	0.47\\
        &Med. Per.&1.67&	0.45&	2.37&	2.26&	2.62&	1.41&	2.57&	1.92&	1.91&	1.19\\
        &1st Quartile&0.72&	0.22&	2.13&	2.26&	2.38&	0.94&	2.34&	1.44&	0.72&	0.95\\
        &3rd Quartile&2.87&	1.56&	2.61&	2.52&	2.86&	1.88&	2.81&	2.16&	2.15&	1.43\\
        &Min. Per.&0.24&	0.22&	0.24&	0.25&	0.24&	0.24&	0.23&	0.24&	0.24&	0.24\\
        &Max. Per.&6.22&	4.46&	8.05&	3.27&	6.67&	4.94&	7.25&	4.79&	5.24&	4.05\\
        \hline
        \multirow{7}{*}{T.L. Pi}&Mean Per.&---	&---&	2.47	&---&	2.61	&---&	2.50	&---	&---	&---\\
        &Std. Dev.&---	&---&	0.65	&---&	0.51	&---&	0.50	&---	&---	&---\\
        &Med. Per.&---	&---&	2.37	&---&	2.62	&---&	2.57	&---	&---	&---\\
        &1st Quartile&---	&---&	2.13	&---&	2.38	&---&	2.34	&---	&---	&---\\
        &3rd Quartile&---	&---&	2.84	&---&	2.86	&---&	2.81	&---	&---	&---\\
        &Min. Per.&---	&---&	0.24	&---&	0.24	&---&	0.23	&---	&---	&---\\
        &Max. Per.&---	&---&	6.87	&---&	7.39	&---&	8.18	&---	&---	&---\\
        \hline
    \end{tabular}
\end{table*}